\documentclass[apj]{emulateapj}
\usepackage{apjfonts}
\newcommand{\tnm}         {\tablenotemark}
\newcommand{\tnt}         {\tablenotetext}
\newcommand{\nogradient}  {\nodata & \nodata}
\begin{document}
\shorttitle{Structure of W3(OH) from 5 cm OH Masers}
\shortauthors{Fish \& Sjouwerman}
\title{Structure of W3(OH) from Very High Spectral
  Resolution Observations of 5 Centimeter OH Masers}
\author{Vincent L.\ Fish\altaffilmark{1} \& Lor\'{a}nt O.\ Sjouwerman}
\affil{National Radio Astronomy Observatory, 1003 Lopezville Road,
  Socorro, NM 87801; vfish@nrao.edu.}
\altaffiltext{1}{Jansky Fellow.}
\begin{abstract}
Recent studies of methanol and ground-state OH masers at very high
spectral resolution have shed new light on small-scale maser
processes.  The nearby source W3(OH), which contains numerous bright
masers in several different transitions, provides an excellent
laboratory for high spectral resolution techniques.  We present a
model of W3(OH) based on EVN observations of the rotationally-excited
6030 and 6035 MHz OH masers taken at 0.024~km\,s$^{-1}$ spectral
resolution.  The 6.0~GHz masers are becoming brighter with time and
show evidence for tangential proper motions.  We confirm the existence
of a region of magnetic field oriented toward the observer to the
southeast and find another such region to the northeast in W3(OH),
near the champagne flow.  The 6.0~GHz masers trace the inner edge of a
counterclockwise rotating torus feature.  Masers at 6030~MHz are
usually a factor of a few weaker than at 6035~MHz but trace the same
material.  Velocity gradients of nearby Zeeman components are much
more closely correlated than in the ground state, likely due to the
smaller spatial separation between Zeeman components.  Hydroxyl maser
peaks at very long baseline interferometric resolution appear to have
structure on scales both smaller than that resolvable as well as on
larger scales.
\end{abstract}

\section{Introduction}
\label{Introduction}

It is only in recent years that astronomers have recognized the
importance of observing astronomical masers at very high spectral
resolution.  Such observations necessarily also require high angular
resolution, as is achieved by very long baseline interferometry
(VLBI), in order to minimize spatial blending of adjacent maser
features, which may be separated by as little as a few
milliarcseconds.  \citet{vlemmings05} observed 22~GHz water masers at
0.03 km~s$^{-1}$ resolution with the Very Long Baseline Array (VLBA)
in order to determine the line profiles, which give clues as to the
saturation, beaming angle, geometry, and physical conditions at the
maser site \citep{nedoluha91,elitzur94,watson02,watson03}.  VLBA
observations of 12~GHz methanol masers at 0.02~km\,s$^{-1}$ resolution
by \citet{moscadelli03} uncovered gradients in the positions of maser
spots as a function of line-of-sight velocity, a phenomenon also
observed in VLBI observations of formaldehyde masers
\citep{hoffman03}.

W3(OH) is a good laboratory for studying maser phenomena.  It contains
a plethora of bright masers without substantial interstellar scatter
broadening.  It is also a close interstellar maser source whose
distance has recently been measured to high accuracy by triangulation
based on maser parallaxes: \citet{xu06} obtain a distance of $1.95 \pm
0.04$~kpc based on methanol maser parallaxes, while
\citet{hachisuka06} derive a distance of $2.04 \pm 0.07$~kpc from
water masers.  (Henceforth, we will assume a distance of 2.00~kpc for
convenience.)  For these reasons, W3(OH) has been a popular target for
VLBI observations of OH masers in the ground-state
\citep{moran68,lo75,mader78,reid80,fouquet82,garciabarreto88,airapetyan89,bloemhof92,masheder94,wright04a,wright04b,fbs06},
4765~MHz \citep{baudry88,baudry91,palmer03}, 6030/6035~MHz
\citep{moran78,desmurs98}, and 13441~MHz \citep{baudry98} transitions,
among numerous observations of OH masers at lower angular resolution.

Recently, \citet[][hereafter FBS]{fbs06} found key differences in the
ground-state transitions of the interstellar OH masers in W3(OH) when
observed at very high spectral resolution.  Velocity gradients appear
to be systematically smaller for 1667~MHz masers than for the other
three transitions (i.e., 1667~MHz spots are more spread out in
position over adjacent channels), matching predictions by
\citet{pavlakis96b}.  Another finding is that 1665~MHz masers
typically have broader line widths than do the other ground-state
transitions.

Contectualizing the empirical differences among the various
transitions of OH may help constrain maser models and thereby provide
additional insight as to the details of the physical processes
involved in OH maser activity.  This is our primary goal in obtaining
similar observations of the 6030 and 6035~MHz masers in W3(OH) at very
high spectral and spatial resolution.  An additional goal is to obtain
very sensitive VLBI observations of the 6.0~GHz lines to determine the
distribution of maser features for comparison with previous and future
epochs of data, both at 6.0~GHz and at other OH maser frequencies.

\section{Observations}
\label{Observations}

The European VLBI Network (EVN) was used to observe the 6030.747 and
6035.092 MHz OH main-line masers in dual circular polarization in
W3(OH).  Eight stations were used: the MkII telescope at
Jodrell Bank, Effelsberg, the Onsala 25~m telescope, Medicina,
Toru\'{n}, Noto, Darnhall, and a single Westerbork antenna.  No usable
data were obtained in left circular polarization (LCP) at 6030 MHz at
the MkII telescope due to a hardware failure.

The observing run (experiment code EF015) lasted for a total of 15
hours on 2006~Jun~11.  W3(OH) was observed in a series of 25-min
blocks each followed by a 5-min scan on the nearby calibrator
J0207+6246.  Seven 10-min scans of the calibrator
3C84 were interspersed for bandpass and single-band delay
calibration.  Sensitivity at the Noto telescope was sufficiently poor
to prevent obtaining detectable fringes even on 3C84, so no data from
Noto were used.

Data were recorded with two-bit sampling of a 500 kHz bandwidth
centered at $v_\mathrm{LSR} = -44$~km~s$^{-1}$ in both frequencies and
both polarizations simultaneously.  For maximum spectral resolution,
the data were correlated in two separate passes, one for each
frequency, at the Joint Institute for VLBI in Europe (JIVE).  Each
baseband channel was divided into 1024 spectral channels, for an
equivalent channel separation of 0.024~km~s$^{-1}$.  A correlator
averaging time of 2~sec was used, which would be expected to produce
negligible ($\sim 1\%$) brightness reduction due to time smearing
at the location of the TW object, 7\arcsec\ to the east
\citep{turner84}.

Data calibration was done in AIPS.  The second calibration table (CL
2) from the EVN user pipeline, which contains a priori amplitude
calibration (setting the absolute flux scale) and parallactic angle
corrections but no fringe-fitting, was used as the starting point for
subsequent data reduction.  The calibrator 3C84 was self-calibrated to
produce a source model.  Several minutes of data on 3C84 were selected
to compute the delay for each antenna using the source model in the
task FRING.  All data on 3C84 were used (again with the source model)
to determine a single complex bandpass correction for each antenna
using BPASS.  The bandpass correction was normalized to the inner 75\%
of the channels.  Bandpass corrections were stable with time; a second
scan-by-scan test run of BPASS on the bandpass-corrected data showed
no apparent variations in the bandpass responses with time.

After running CVEL to apply the bandpass correction to the data and
correct the velocities to the local standard of rest, a single channel
in a single circular polarization containing bright maser emission in
W3(OH) was SPLIT out.  The data for this channel were self-calibrated.
The resulting calibration was applied to all spectral channels in both
circular polarizations of the same frequency, although 6030 MHz and
6035 MHz data were calibrated independently.  A large, lower spatial
resolution and lower spectral resolution image was created using IMAGR
with fields centered near W3(OH) itself and the TW object in order to
identify regions of maser emission.  Higher resolution images (2~mas
pixels) in the subfields with detected maser emission were produced at
full spectral resolution.  The resulting natural-weighted synthesized
beam width was approximately 7~mas.  In each spectral channel,
detected maser spots were fitted with elliptical Gaussians using a
task based on JMFIT.  In cases where the separation between two nearby
maser spots is comparable to the synthesized beam width, elliptical
Gaussians were fit simultaneously to both features.

The positional offset from the calibrator J0207+6246 was derived by
fringe-fitting the bright reference maser spot used for
self-calibration.  The fringe fit solution (phase and fringe rate) was
applied to the calibrator.  Time ranges of poor phase stability were
flagged and the remaining data were imaged.  The apparent position of
the calibrator was compared to the assumed position from the VLBA
Calibrator
List\footnote{\url{http://www.vlba.nrao.edu/astro/calib/vlbaCalib.txt}},
derived from the Second VLBA Calibrator Survey
\citep[VCS2,][]{fomalont03}, to determine the absolute position of the
reference maser feature.

Blank sky noise ranges from 11 to 16~mJy\,beam$^{-1}$, depending on
frequency, polarization, and spectral channel.  These values are in
line with theoretical noise levels from the EVN
Calculator\footnote{\url{http://www.evlbi.org/cgi-bin/EVNcalc}},
suggesting that the absolute flux scale is reasonable.  However,
dynamic range limitations dominate in many channels due to very bright
emission, with noise levels exceeding 0.5~Jy\,beam$^{-1}$ in some
channels.  Sidelobe contamination from the brightest spots is severe
and may have contaminated the spectra of nearby maser spots centered
near the same velocity.

\section{Results}
\label{Results}

\begin{figure}[t]
\resizebox{\hsize}{!}{\includegraphics{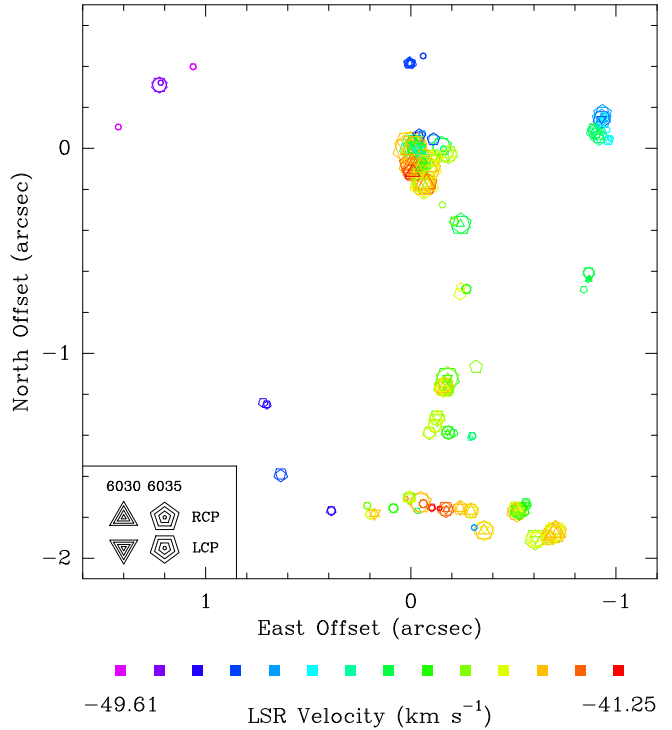}}
\caption{Map of detected 6030 and 6035~MHz maser emission in W3(OH).
  Symbol size is proportional to the logarithm of the peak brightness,
  with symbol sizes indicating 0.1, 1, 10, and 100~Jy~beam$^{-1}$
  shown in the lower left.  LSR velocity is shown in color.  LCP and
  RCP masers are each shown as open symbols for clarity.  The center
  corresponds to $02^\mathrm{h}27^\mathrm{m}03\fs8343,
  +61\degr52\arcmin25\farcs300$ (J2000), with errors as detailed in
  \S\ref{previous}.
\label{fig-map}
}
\end{figure}

\begin{figure}[t]
\resizebox{\hsize}{!}{\includegraphics{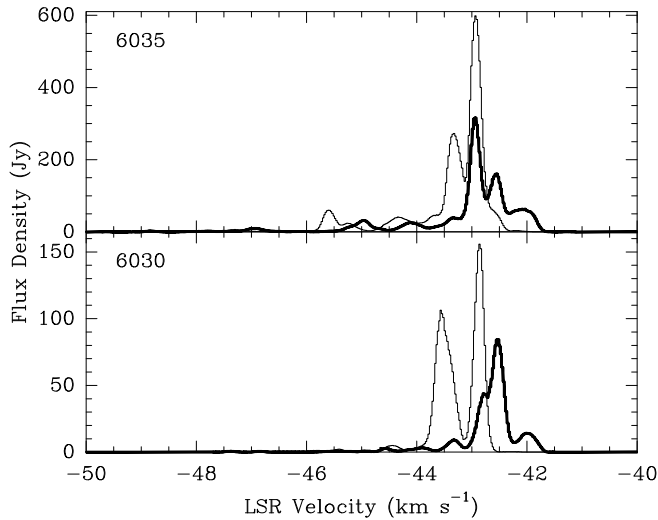}}
\caption{Spectra of maser emission in W3(OH) at 6035 (top)
  and 6030 MHz (bottom).  RCP emission is shown in bold and LCP
  emission in normal weight.
\label{fig-spectra}
}
\end{figure}

The maser components we identify are listed in Table \ref{tab-spots}
and shown in Figure~\ref{fig-map}.  The brightest masers, including
the reference features at 6030 and 6035~MHz, are located in the
cluster near the origin.  No spots were seen near the TW object.
Spectra of the maser emission, obtained by summing the flux density of
all maser spots detected, are shown in Figure \ref{fig-spectra}.
Overall, we find 90 spots at 6030 MHz and 202 spots at 6035 MHz
associated with W3(OH).  The weakest masers we detected have
brightnesses of $\sim 100$~mJy\,beam$^{-1}$.

\subsection{Comparison with Previous Studies}
\label{previous}

There have been two previous VLBI studies of the 6.0~GHz masers in
W3(OH).  \citet{moran78} detected 12 Zeeman pairs at 6035 MHz with a
three-element interferometer.  \citet[][hereafter D98]{desmurs98}
observed both main-line 6.0~GHz transitions with three elements of the
EVN (Effelsberg, Mk~II, and Medicina).  Present observations exceed
both previous observations in sensitivity and spectral resolution and
produce a better sampling of the $uv$-plane.

The reference feature at 6030~MHz (spot 168, 82~Jy) is the brightest
right circular polarized (RCP) feature in that transition and was
chosen in preference to the brighter LCP feature in order that data
from the Mk~II telescope could be calibrated as well.  The reference
feature at 6035 MHz (spot 226, flux density 579~Jy LCP) is the
brightest maser feature in W3(OH).  We derive a position of
$02^\mathrm{h}27^\mathrm{m}03\fs8249, +61\degr52\arcmin25\farcs217
(\mathrm{J}2000)$ for the reference spot at 6030~MHz and
$02^\mathrm{h}27^\mathrm{m}03\fs8343, +61\degr52\arcmin25\farcs300$
for the reference spot at 6035~MHz.  Formal errors are approximately
1.5~mas, but this is likely an underestimate due to the poor quality
of the images of the calibrator produced after fringe-fitting on the
W3(OH) reference spots.  We estimate that actual errors are
approximately 10~mas, based on the offset needed to align overlapping
Zeeman pairs at 6030 and 6035~MHz (Table~\ref{tab-zeeman}), as
discussed in \S \ref{zeeman}.  The positions in Table~\ref{tab-spots}
reflect this offset.  Our positions agree with those of
\citet{etoka05} to within 50~mas and with those of D98 to within
180~mas.  For comparison, a typical maser motion of 3~km\,s$^{-1}$
\citep{bloemhof92} would produce a net motion of 4~mas between the
epoch of observations of D98 and this work.

\begin{figure*}[t]
\resizebox{\hsize}{!}{\includegraphics{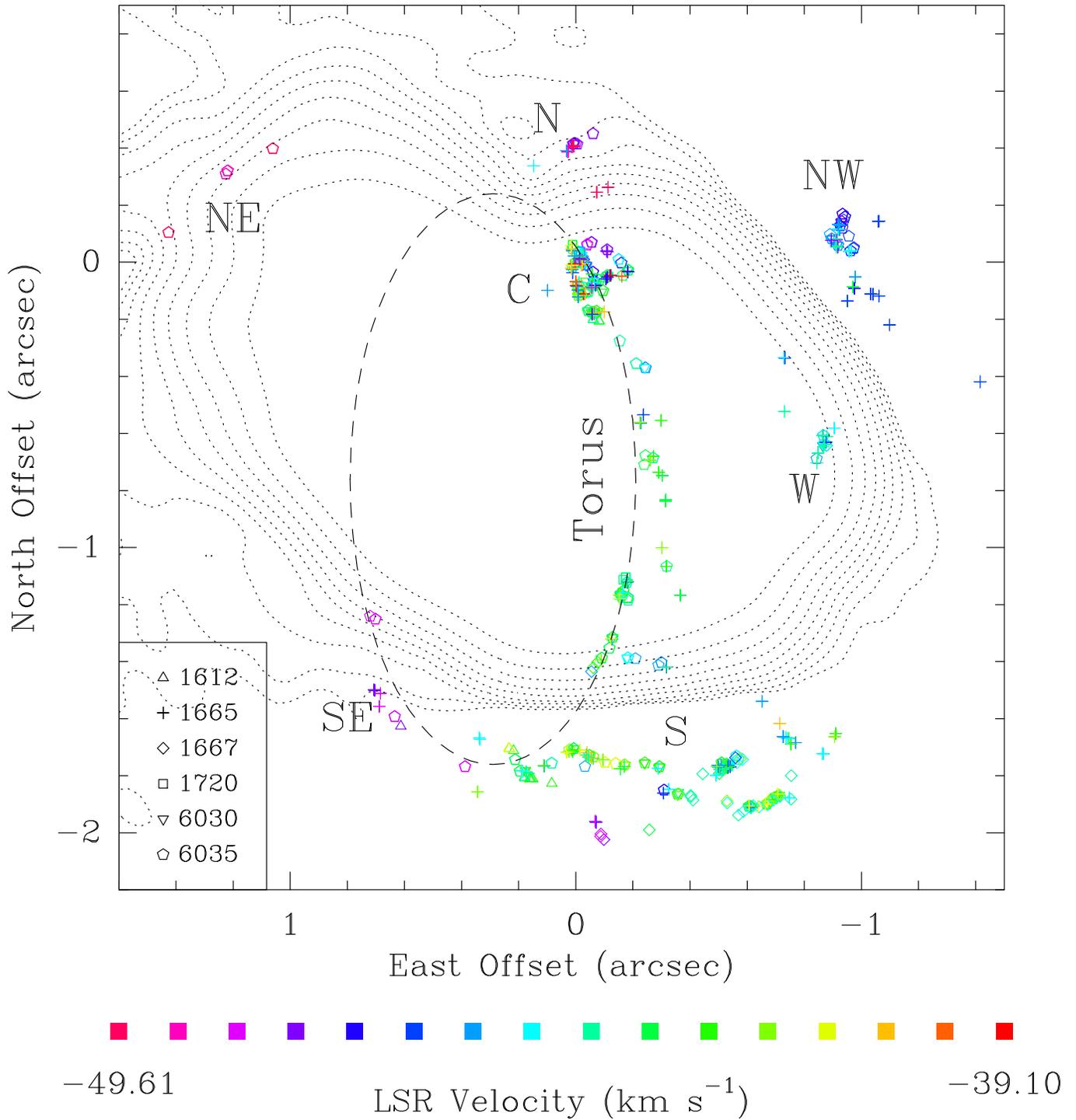}}
\caption{Map of 6030 and 6035~MHz masers as well as ground-state
  masers (from FBS), superposed atop 8.4~GHz continuum emission from
  the VLA archive \citep[experiment code AR363, see][]{wilner99}.
  Velocities are not corrected for Zeeman splitting (see
  Fig.~\ref{fig-zeeman-vel}), which can be large (several kilometers
  per second) and unknown (due to unpaired Zeeman components) at 1665
  and 1667~MHz.  Dotted contours are shown at 4, 8, 16, \ldots times
  the rms noise of 11.6~$\mu$Jy\,beam$^{-1}$.  Key maser groups and a
  potential torus feature (discussed in detail in \S \ref{structure})
  as traced by the 6.0~GHz masers are indicated.
\label{fig-map-all}
}
\end{figure*}

It is not possible to align the ground-state and excited-state data to
milliarcsecond accuracy, since the ground-state data were not phase
referenced in FBS.  Nevertheless, an approximate registration can be
obtained from the similarity in distribution of the ground-state and
6.0~GHz data (Fig.~\ref{fig-map-all}).  Masers at 6035~MHz are found
both in regions where numerous ground-state masers occur as well as in
regions where no ground-state emission is detected.  However, 6035~MHz
masers in different regions are seen in approximate superposition with
each of the four ground-state transitions.

\begin{figure}[t]
\resizebox{\hsize}{!}{\includegraphics{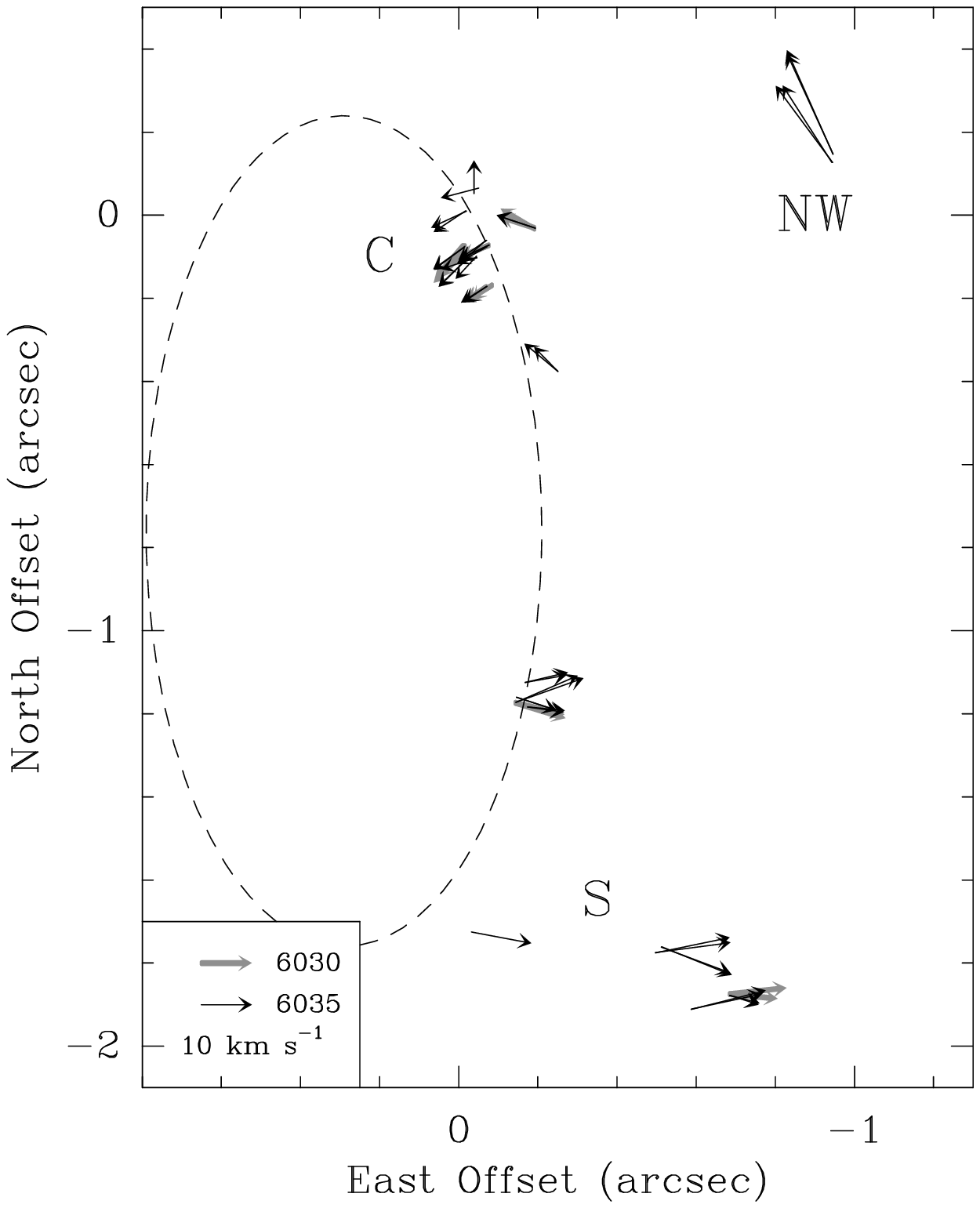}}
\caption{Motions of 6030 and 6035~MHz masers between the D98 epoch
  (1994 May 20--23) and the present observations.  Vector length is
  proportional to velocity, with a 10~km\,s$^{-1}$ motion indicated in
  the inset.  The data are consistent with the addition of a single
  constant vector to the motions in each of the 6030 and 6035~MHz
  transitions.  These constant vectors have been chosen to minimize
  the rms motions.  The motions are not consistent with those obtained
  in the ground-state by \citet{bloemhof92}.
\label{fig-motions}
}
\end{figure}

We can identify most of the features of D98 in our present data.
Their reference feature at 6035~MHz (component A) corresponds to our
spots 91 and 93, while their reference feature at 6030~MHz (component
A$^\prime$) corresponds to our spots 168 and 169.  We note, as do
\citet{etoka05}, a systemic shift in LSR velocity of approximately
0.4~km~s$^{-1}$ in the D98 data.  Figure~\ref{fig-motions} shows the
motions derived from comparing our data with that of D98.  Owing to
differences in the positions of the reference features obtained in the
two epochs, the absolute alignment of the maps of the two epochs of
data is not known.  Thus, the data are consistent with any set of
motions differing from Figure~\ref{fig-motions} by the addition of a
constant vector to all motions in each frequency (different at 6030
and 6035~MHz).  We have chosen these constant vectors to minimize the
root-mean-square (rms) motion, which is 10~km\,s$^{-1}$.  The motions
appear to be indicating fast counterclockwise rotation, in contrast
with the 3 to 5~km\,s$^{-1}$ expansion obtained from ground-state
motions \citep{bloemhof92}.  \citet{wright04a} note a possible
component of rotation based on 1665~MHz proper motions, although they
interpret their motions as a \emph{clockwise} rotation (i.e., north to
south along the face of the continuum emission).  In \S
\ref{structure} we argue that if the masers in the SE cluster in
Figure~\ref{fig-map-all} are part of this structure, rotation is in
the counterclockwise sense.  We do not discount the possibility that
we have misidentified several maser spots between the two epochs
partially due to spatial blending of maser spots in both epochs,
although this would add random noise rather than a systematic bias to
the derived motions.

\subsection{Variability}
\label{variability}

Of the 292 detected features, 208 (71\%) are brighter than
400~mJy\,beam$^{-1}$, the brightness of the weakest spot in D98, who
identified 82 maser features with three EVN antennas.  For comparison,
203 features (80\%) brighter than 400~mJy\,beam$^{-1}$ were seen in
the four ground-state lines in FBS.  It is surprising that the number
of 6.0~GHz masers above this flux threshold exceeds that of
ground-state masers, and even more surprising that the brightest
6.0~GHz maser (over 575~Jy\,beam$^{-1}$) is far brighter than the
brightest ground-state maser ($< 100$~Jy\,beam$^{-1}$, from FBS).
Surveys of other regions at lower resolution indicate that 6035~MHz
masers are on average much weaker and less common by a factor of three
than 1665~MHz masers \citep{caswell01,caswell03}.

\begin{figure}[t]
\resizebox{\hsize}{!}{\includegraphics{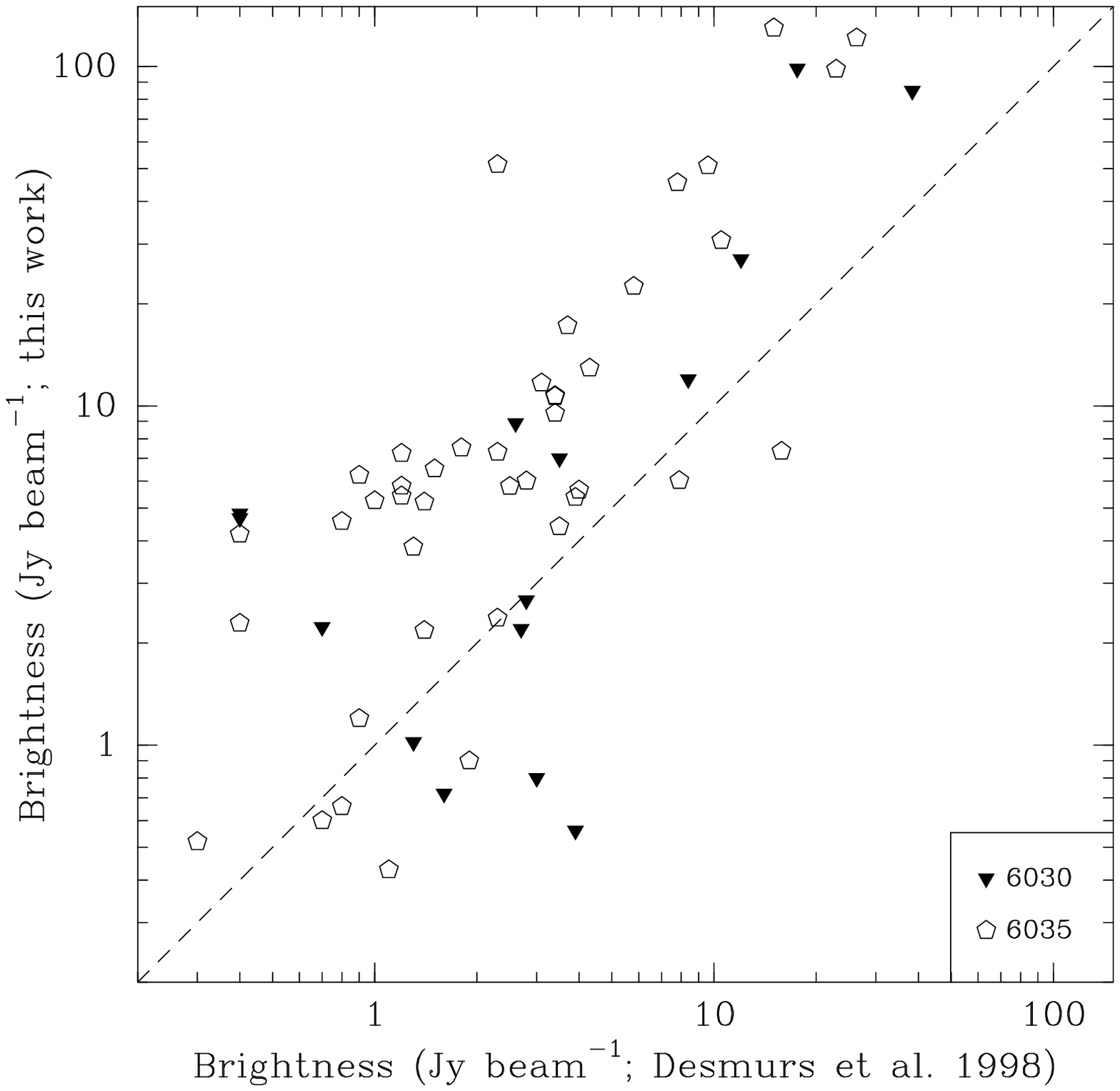}}
\caption{Variability of masers identified in both D98 and the present
  epoch.  The dashed line indicates no change.  Most masers have
  become brighter since the D98 epoch, assuming both flux scales are
  correct.
\label{fig-variability}
}
\end{figure}

However, the two sets of observations are not cotemporal.  Indeed, the
6.0~GHz masers in W3(OH) are becoming brighter with time.
Figure~\ref{fig-variability} shows the change in brightness for masers
identified in both D98 and the current epoch (the same sample used to
obtain proper motions).  Many of the redetected masers are now
approximately four times as bright as in the D98 epoch.  As comparison
of Figure~\ref{fig-spectra} with the D98 Effelsberg autocorrelation
spectra indicates, this is a real effect.  Most masers that have
become weaker are spatially blended with other maser spots in one or
both epochs.  This brightening is observed among 6.0~GHz masers in all
regions of W3(OH), not just (for instance) the cluster near the origin
(C).  The D98 total power spectra also indicate a general brightening
compared to the \cite{moran78} data, especially in the brightest LCP
feature, although some individual features detected by \citet{moran78}
are less bright in the D98 data.

\subsection{Zeeman Associations}
\label{zeeman}

Zeeman pairs are listed in Table~\ref{tab-zeeman}.  Listed positions
are the arithmetic mean of the locations of the LCP and RCP maser
spots in each Zeeman pair.  We identify a total of 117 Zeeman pairs,
35 in the 6030 MHz transition and 82 in the 6035 MHz transition.
These include 28 Zeeman associations consisting of a Zeeman pair at
each of 6030 and 6035 MHz.  The difference in systemic velocity
between the 6030 and 6035~MHz pairs is less than or equal to
0.10~km\,s$^{-1}$ in 22 of the 28 full Zeeman quartets.  The maximum
velocity difference, 0.21~km\,s$^{-1}$ in association 7, is still
within a typical line width, and the slightly different positions and
velocity gradients (in excellent agreement in magnitude and position
angle between components of a Zeeman pair in one transition but not
between components of different transitions) argue that the 6030 and
6035~MHz masers each trace different (though closely associated)
material.  Magnetic field values agree to within 1.0~mG or better in
24 of the 28 Zeeman associations.  Including individual spots in one
transition consistent with the magnetic field and central velocity at
the same location in the other transition (parenthesized in
Table~\ref{tab-zeeman}), 84\% of maser spots are found in a Zeeman
association.

\begin{figure}[t]
\resizebox{\hsize}{!}{\includegraphics{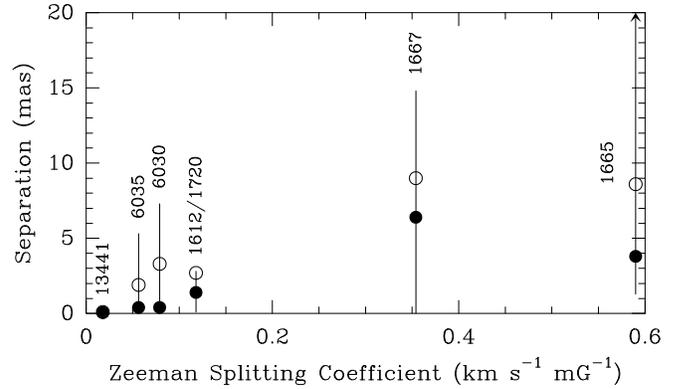}}
\caption{Separation between components of a Zeeman pair vs.\ Zeeman
  splitting coefficient.  Lines indicate the range of separations,
  with median and 90\% values indicated as filled and open circles,
  respectively.  Satellite-line ground-state data are combined due to
  the small number of spots at 1612 and 1720 MHz.  Zeeman pair
  separations increase with increasing Zeeman splitting coefficient.
  Ground-state data are from FBS; 13441~MHz data are from
  \citet{baudry98}.  
\label{fig-seps-freq}
}
\end{figure}

The median separation of components of a 6.0~GHz Zeeman pair is
0.4~mas (0.8~AU), with 90\% of pairs having separations within 2.0~mas
(4.0~AU).  The 6.0~GHz Zeeman pairs have smaller separations than
those in the ground state, as shown in Figure \ref{fig-seps-freq}.
The median separation between centers of 6030 MHz and 6035 MHz Zeeman
pairs in a Zeeman association is 0.9~mas (1.8~AU), with 90\% within
3.0~mas (6.0~AU).

\begin{figure}[t]
\resizebox{\hsize}{!}{\includegraphics{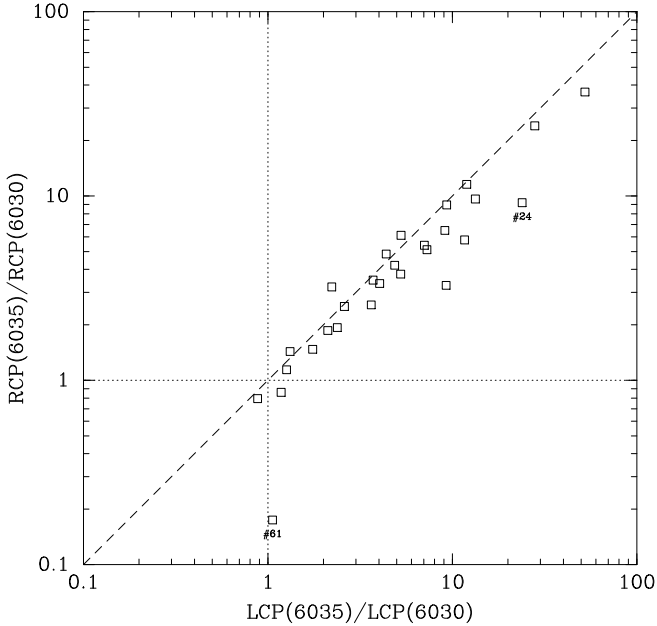}}
\caption{Plot of ratios of 6035 to 6030~MHz peak brightnesses of RCP
  vs.\ LCP spots in the 28 Zeeman associations consisting of a Zeeman
  pair at each frequency.  Most points fall near the dashed line,
  which corresponds to equal amplification of the 6035~MHz Zeeman pair
  components compared to their 6030~MHz counterparts.  The slight
  offset below this line is likely due to a small calibration error in
  the 6035~MHz RCP data.  The dotted lines divide regions in which the
  6030 and 6035~MHz masers dominate in brightness; in nearly every
  Zeeman association the 6035~MHz masers are brighter, usually by a
  factor of several.  Zeeman associations 24 and 61, discussed in \S
  \ref{zeeman}, are labelled.
\label{fig-zeeman-ratios}
}
\end{figure}

\begin{figure}[t]
\resizebox{\hsize}{!}{\includegraphics{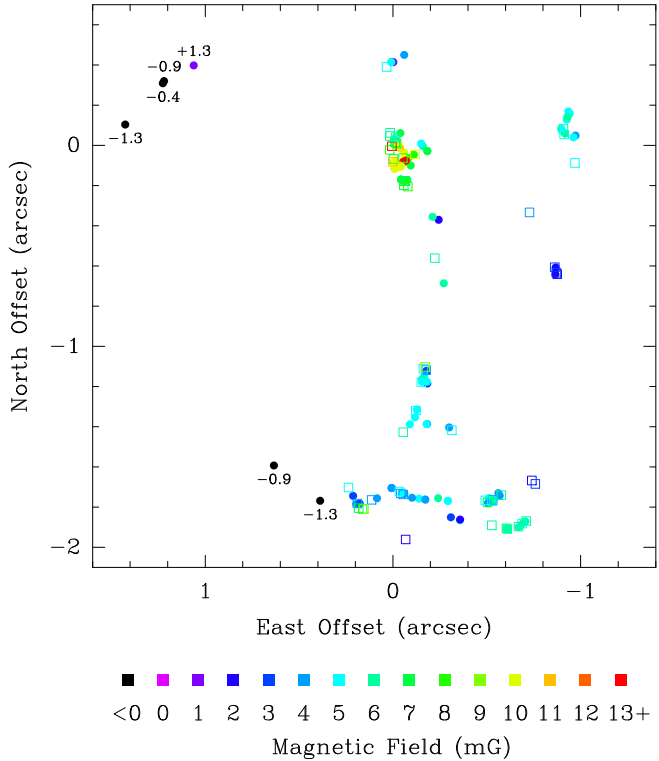}}
\caption{Magnetic field as determined from OH maser Zeeman splitting.
  Zeeman pairs at 6.0~GHz (this work) are shown as filled circles, and
  ground-state Zeeman pairs (FBS) are shown as open squares.  Colored
  symbols indicate the magnetic field strength (specifically, $\lfloor
  B \rfloor$ for positive magnetic fields) in increments of 1~mG.
  Negative values, indicating a line-of-sight magnetic field oriented
  toward the observer, are shown in black.  Labels denote magnetic
  field values for Zeeman pairs 84--89.
\label{fig-zeeman-map}
}
\end{figure}

Masers at 6030 and 6035~MHz in Zeeman associations appear to be
sampling the same material.  In addition to being located at
approximately the same location and sampling the same magnetic field,
Zeeman pairs at 6035~MHz are brighter than, and preserve approximately
the same brightness ratios as, their counterpart at 6030~MHz in Zeeman
associations (Fig.~\ref{fig-zeeman-ratios}).  Comparison of the VLBI
maser spectra with Effelsberg auto-correlation data suggest that the
slight offset from equal additional amplification in both
polarizations at 6035~MHz compared to 6030~MHz is due to a larger
fractional unrecovered flux at 6035~MHz RCP compared to the other
polarization and transition.  The largest outlier data point, Zeeman
association 61 (Table~\ref{tab-zeeman}) is located near the origin in
Figure~\ref{fig-zeeman-map}, where the magnetic field changes rapidly
on a small spatial scale, differing by over 4~mG between the 6030 and
6035~MHz Zeeman pairs.  It is probable that the 6030 and 6035~MHz
pairs do not sample the same material in this association, given the
discrepant magnetic field values obtained from their Zeeman splitting.

Zeeman association 24 also has a large difference between 6030 and
6035~MHz estimates of the magnetic field.  We note that the spatial
separation of the 6030~MHz Zeeman pair is the largest of any 6.0~GHz
Zeeman pair (7.3~mas).  The RCP component of this pair (spot 78) is
significantly closer to the 6035~MHz emission and has an LSR velocity
consistent with the central velocity and magnetic field derived at
6035~MHz.  It is likely that spots 75 and 78 are not a true Zeeman
pair but instead ``Zeeman cousins'' \citep[opposite circular
polarization components of two different Zeeman pairs in the same
cluster of masers, as defined in][]{fish06}.

Figure~\ref{fig-zeeman-map} shows a map of the magnetic field
strengths as determined from 6030 and 6035~MHz Zeeman pairs.  Magnetic
field strengths are broadly consistent with previous observations,
with the largest magnetic field strengths found near the origin.  We
do not find a systemic bias for higher magnetic field strengths in
either transition compared to the other.  Of the 28 overlapping
associations of Zeeman pairs at both 6030 and 6035~MHz, 15 indicate
larger magnetic field strengths at 6030~MHz and 13 at 6035~MHz.  A
similar conclusion was reached by D98.

\begin{figure*}[t]
\resizebox{\hsize}{!}{\includegraphics[angle=-90]{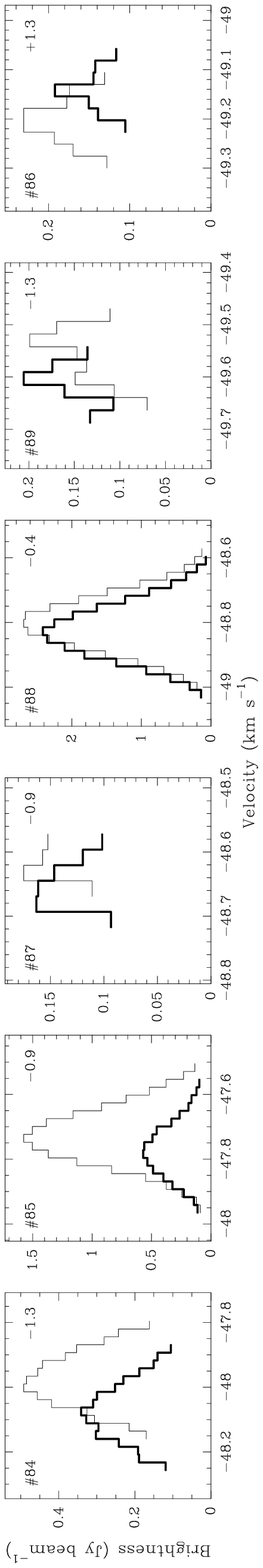}}
\caption{Spectra of Zeeman pairs indicating a negative magnetic field.
  RCP emission is shown in bold and LCP in normal weight.  In the
  first five plots, the LCP emission is shifted to higher (less
  negative) velocity than the RCP emission, indicating a line-of-sight
  magnetic field oriented toward the observer.  The Zeeman pair number
  (Table~\ref{tab-zeeman}) and magnetic field (in mG) are indicated in
  the upper left and right of each plot, respectively.  Zeeman pairs
  84 and 85 are the easternmost Zeeman pairs in the southeast (SE)
  maser group of W3(OH). Pairs 87, 88, and 89 are the easternmost
  pairs in the northeast cluster (NE).  The last plot, pair 86,
  indicates a positive magnetic field.  This pair is the westernmost
  of the four Zeeman pairs in the northeastern cluster.
\label{fig-minus-B}
}
\end{figure*}

Magnetic field values are nearly everywhere positive (i.e., in the
hemisphere pointing away from the observer).  However, we identify a
total of five Zeeman pairs, all in the 6035~MHz transition, with the
opposite line-of-sight field direction.  The velocity range of the
spots in these Zeeman pairs is very clean, so it is unlikely that
sidelobe contamination has caused us to mistake their velocities.
Spectra of these Zeeman pairs is shown in Figure~\ref{fig-minus-B}.
While several of the features in individual Zeeman pairs are weak,
taken together the sample provides strong evidence for a line-of-sight
magnetic field reversal in the source.  Several of these pairs are
located farther east than previous detections at lower (more
negative) LSR velocity.  \citet{moran78} claim the detection of a
Zeeman pair with a magnetic field of $+2$~mG in the northeast, but the
separation between the two spots is $275 \pm 100$~mas, making it
likely that the two spots are components of two different Zeeman pairs
instead.  Neither D98 nor \citet{etoka05} note 6035~MHz masers at the
velocities or locations of the negative magnetic field Zeeman pairs,
although the Effelsberg autocorrelation data in Figure~1 of D98 shows
weak evidence of a feature near $-49$~km\,s$^{-1}$ after correcting
for the previously noted velocity shift.  Three or four (depending on
the exact velocity shift of the D98 data) of the five Zeeman pairs
indicating a negative magnetic field in our data, including the only
pair brighter than 1~Jy\,beam$^{-1}$ in both circular polarizations,
lie outside the velocity range imaged in D98.  \citet{fisheff06} also
found evidence for negative magnetic fields at a velocity outside our
observed range ($-70$~km\,s$^{-1}$) in the direction of W3(OH), though
the association of these masers with W3(OH) itself rather than another
source within the $130\arcsec$ Effelsberg beam has not been confirmed
interferometrically.

Comparison of the 6.0~GHz Zeeman data with ground-state data (FBS)
shows general agreement between magnetic field values.  Where
ground-state masers and excited-state masers overlap approximately,
magnetic field strengths usually agree to within 1 to 2~mG, the
principal exceptions being in the cluster near origin in
Figure~\ref{fig-zeeman-map}, where the magnetic field varies greatly
over small spatial scales.  In the extreme southeast of the
distribution of masers in W3(OH), \citet{wright04a,wright04b} found
ground-state masers at velocities suggestive of a negative magnetic
field.  Our 6035~MHz data are consistent with this conclusion.  As for
the 13441~MHz masers, we note that that aligning the Zeeman pairs with
a magnetic field of 7.7~mG in \citet{baudry98} near Zeeman pair 49
(7.7~mG) in Table~\ref{tab-zeeman} also produces excellent general
agreement with the other Zeeman pairs in the \citeauthor{baudry98}
data.  In particular, magnetic field strengths of 10.0~mG or greater
in the \citet{baudry98} data match up near Zeeman associations 58, 67,
and 68 in Table~\ref{tab-zeeman}.  Thus, while the 13441~MHz OH masers
preferentially trace material at higher magnetic field strengths than
do the 6.0~GHz masers in W3(OH), the same set of physical conditions
can be conducive to co-propagation of 6030, 6035, and 13441~MHz
masers, a point supported both by theory and other observations
\citep[e.g.,][]{gray01,caswell04}.  A 10.3~mG Zeeman pair at 13434~MHz
centered at $-42.21$~km\,s$^{-1}$ likely comes from the same area
\citep{gusten94,fish05}.

A reanalysis of data from FBS shows no evidence of ground-state masers
in the regions where 6035~MHz masers indicate a negative magnetic
field.  There are no ground-state masers in these regions brighter
than $5\,\sigma$.  Blank sky rms noise levels are
20--30~mJy\,beam$^{-1}$, depending on transition, polarization, and
spectral channel.  Small velocity ranges containing very bright
emission are dynamic-range limited and do not achieve theoretical
blank-sky noise levels.  Our usable bandwidth extends to
$v_\mathrm{LSR} = -49.15$~km\,s$^{-1}$ at 1720~MHz and
$-49.51$~km\,s$^{-1}$ at 1612~MHz, with 1665 and 1667~MHz edge
velocities intermediate to these two values.

\subsection{Line Profiles}
\label{profiles}

For each maser line profile that is (near-)Gaussian in shape or has a
clear (near-)Gaussian peak, we have fitted a Gaussian to obtain the
FWHM (full width at half maximum) of the peak of the profile.  We
obtain a FWHM line width for 169 of the 292 masers in our sample.
Fits were done using brightnesses rather than flux densities since
nearly all masers are pointlike at our angular resolution
(\S\ref{deconvolved}).  Masers for which we do not obtain a measure
of the FWHM have been excluded for one of several reasons.  Some
masers are too weak or have been detected in too few spectral channels
to obtain good constraints on the fit parameters.  Other masers have
irregular or broad asymmetric spectral profiles or appear to consist
of two or more Gaussian components of similar amplitude blended
together.  Some masers have profiles that are too heavily contaminated
by emission from other nearby masers to obtain a good estimate of the
line width.

\begin{figure}[t]
\resizebox{\hsize}{!}{\includegraphics{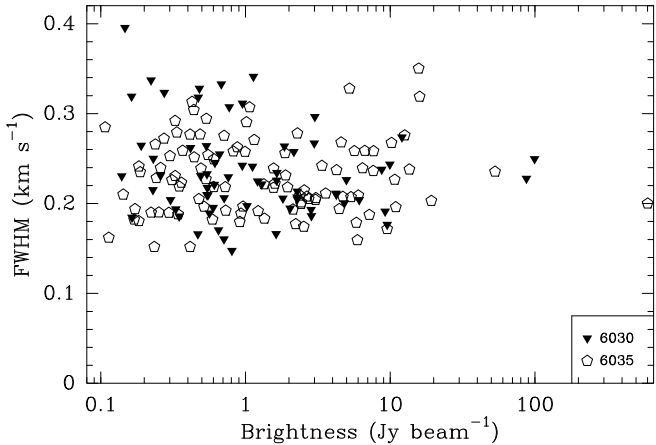}}
\caption{Distribution of FWHM maser line widths as a function of
  fitted brightness.  As in the ground-state (FBS), there does not
  exist a strong correlation between line width and brightness.
\label{fig-fwhm}
}
\end{figure}

The range of FWHM line widths of the 6030 and 6035~MHz masers in
W3(OH) is 0.15--0.40~km\,s$^{-1}$, almost exactly the range of line
widths seen in the ground state (FBS).  Bright masers do not in
general appear broader than weak masers (Fig.~\ref{fig-fwhm}),
consistent with results of \citet{baudry97} and D98.  The mean and
sample standard deviations of the FWHM line widths are $0.235 \pm
0.050$~km\,s$^{-1}$ at 6030~MHz and $0.229 \pm 0.041$~km\,s$^{-1}$ at
6035~MHz.  These values are in excellent agreement with those obtained
at 1665~MHz (FBS).  For comparison, the 13441~MHz masers have FWHM
line widths of $0.24 \pm 0.02$~km\,s$^{-1}$ \citep{baudry98},
consistent with the mean of 6.0~GHz masers though with less scatter.
The distributions of line widths of 6030 and 6035~MHz masers are
consistent with being identical with each other, a conclusion
supported by \citet{baudry97} but not D98, who find broader 6035~MHz
line widths.  It is not clear where this discrepancy arises, although
possibilities include the higher spectral resolution of our data as
compared to that of D98 as well as our exclusion of distinctly
non-Gaussian line shapes from our analysis.  Our better $uv$-coverage
and sensitivity may also provide less sidelobe contamination.  In any
case, blending of nearby maser components might be expected to broaden
line widths more at 6035~MHz than at 6030~MHz due to the larger number
of maser spots in the former transition.

Even maser spectra that look essentially Gaussian are normally not
well fit with a single Gaussian profile.  Figure~\ref{fig-quads} shows
spectra of selected Zeeman associations.  Line shapes range from
near-Gaussian to asymmetric.  In one case (Zeeman associations 39 and
41), double-peaked profiles at 6035~MHz are clearly indicative of
spatial blending of two distinct maser spots.  Other cases are less
clear; see \S\S \ref{deconvolved} and \ref{maser} for a more complete
discussion.

\subsection{Deconvolved Spot Sizes}
\label{deconvolved}

Most spots are unresolved with the EVN.  Maser features with peak
brightness greater than 1~Jy\,beam$^{-1}$, with a clearly detected
peak in the spectral domain, and well fit spatially by a single
elliptical Gaussian are detected in a total of over 2000
spot-channels.  Of these, 83\% have a nominal deconvolved spot size of
zero, and 90\% are consistent with a zero minimum deconvolved spot
size.  Thus, in the vast majority of cases, our spatial resolution is
insufficient to determine the actual maser spot size.

\begin{figure*}
\resizebox{\hsize}{!}{\plottwo{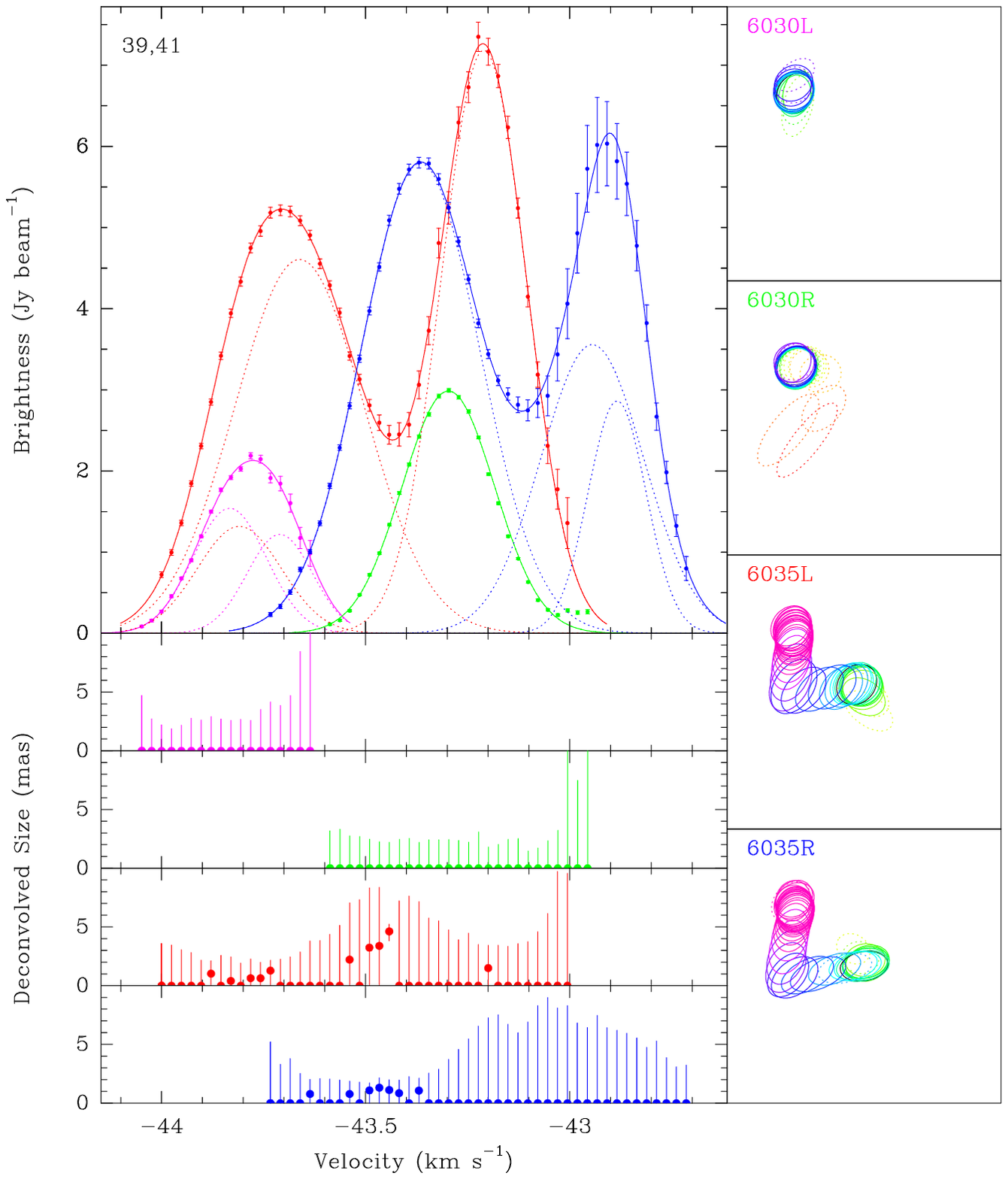}{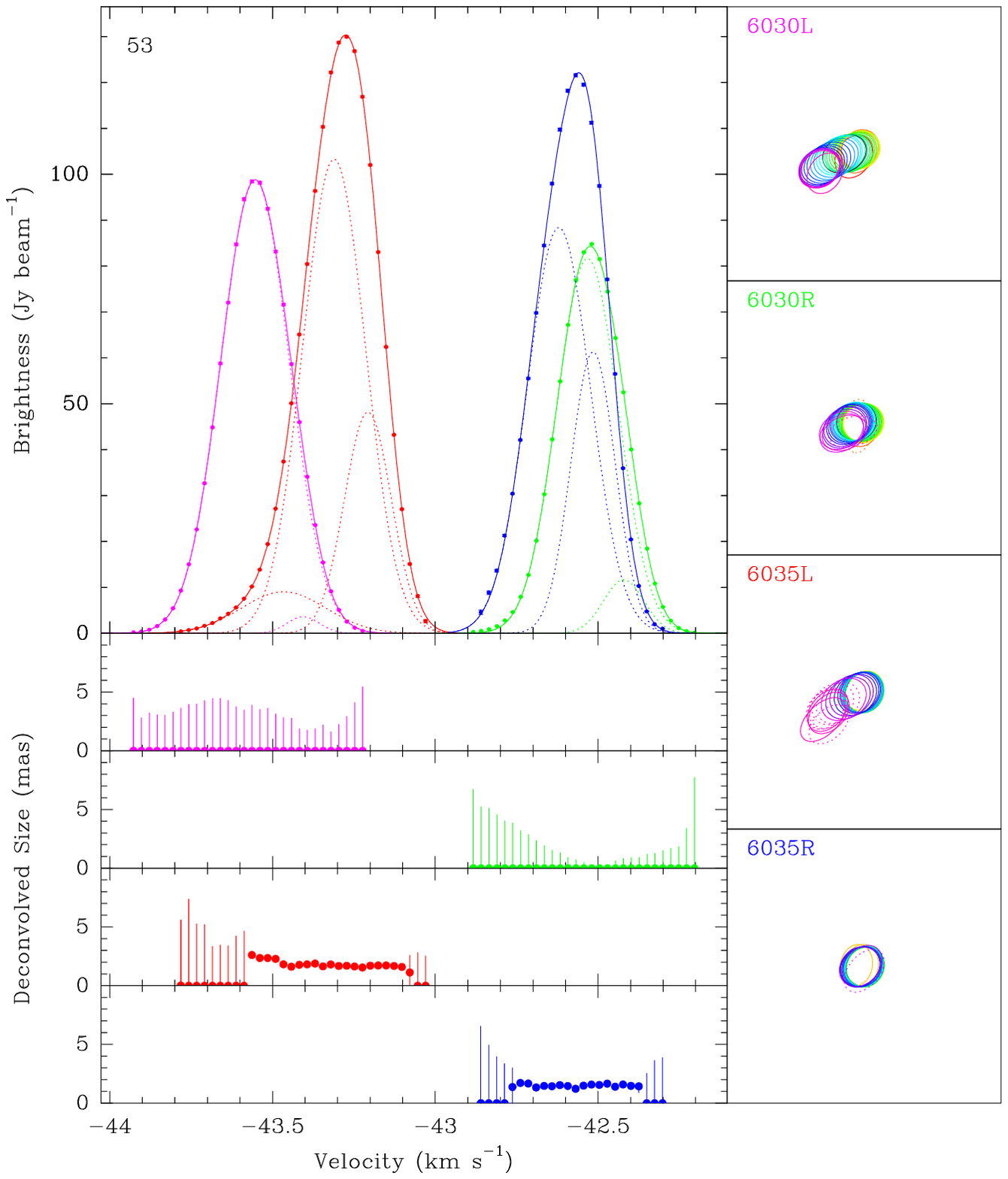}}
\resizebox{\hsize}{!}{\plottwo{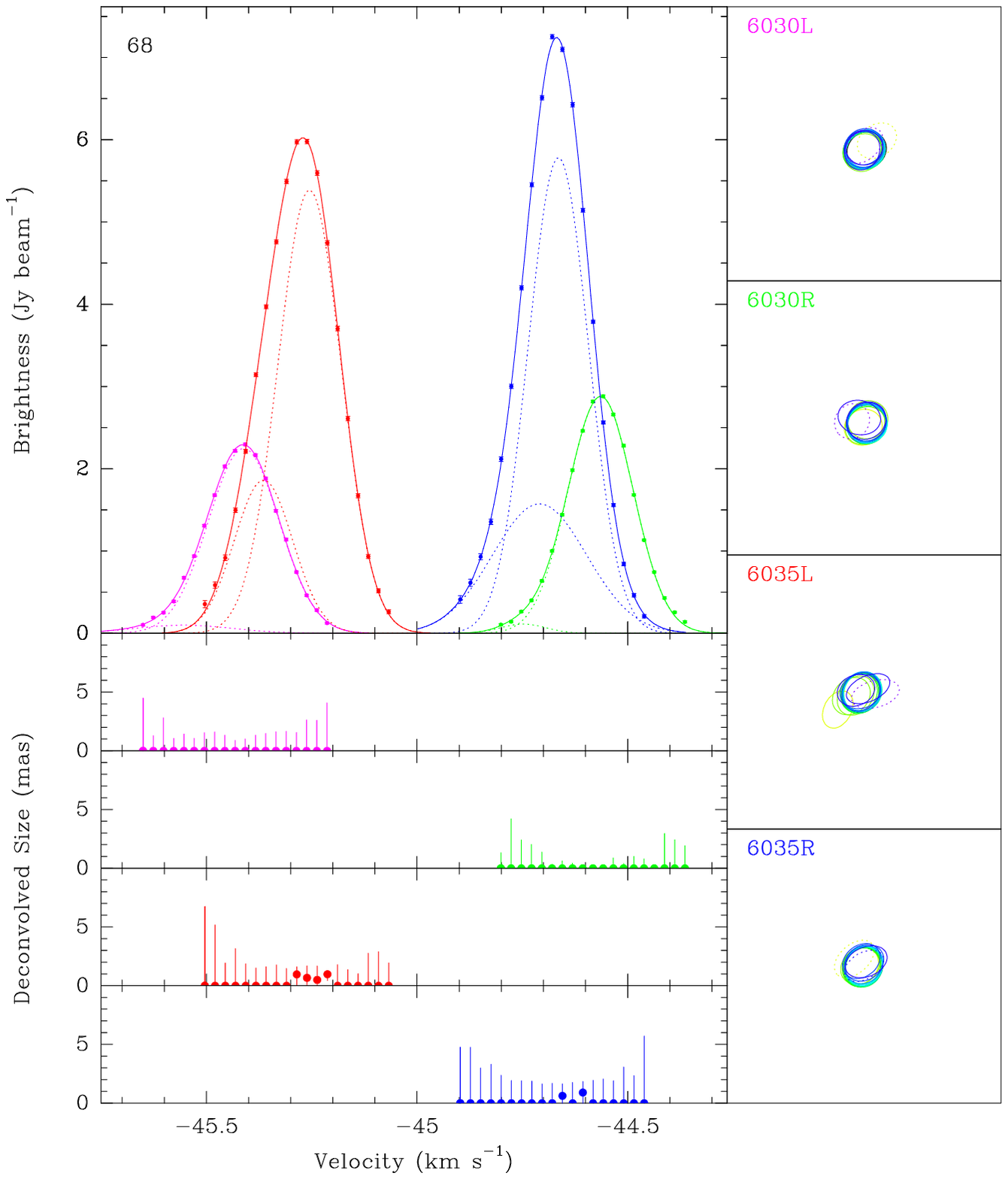}{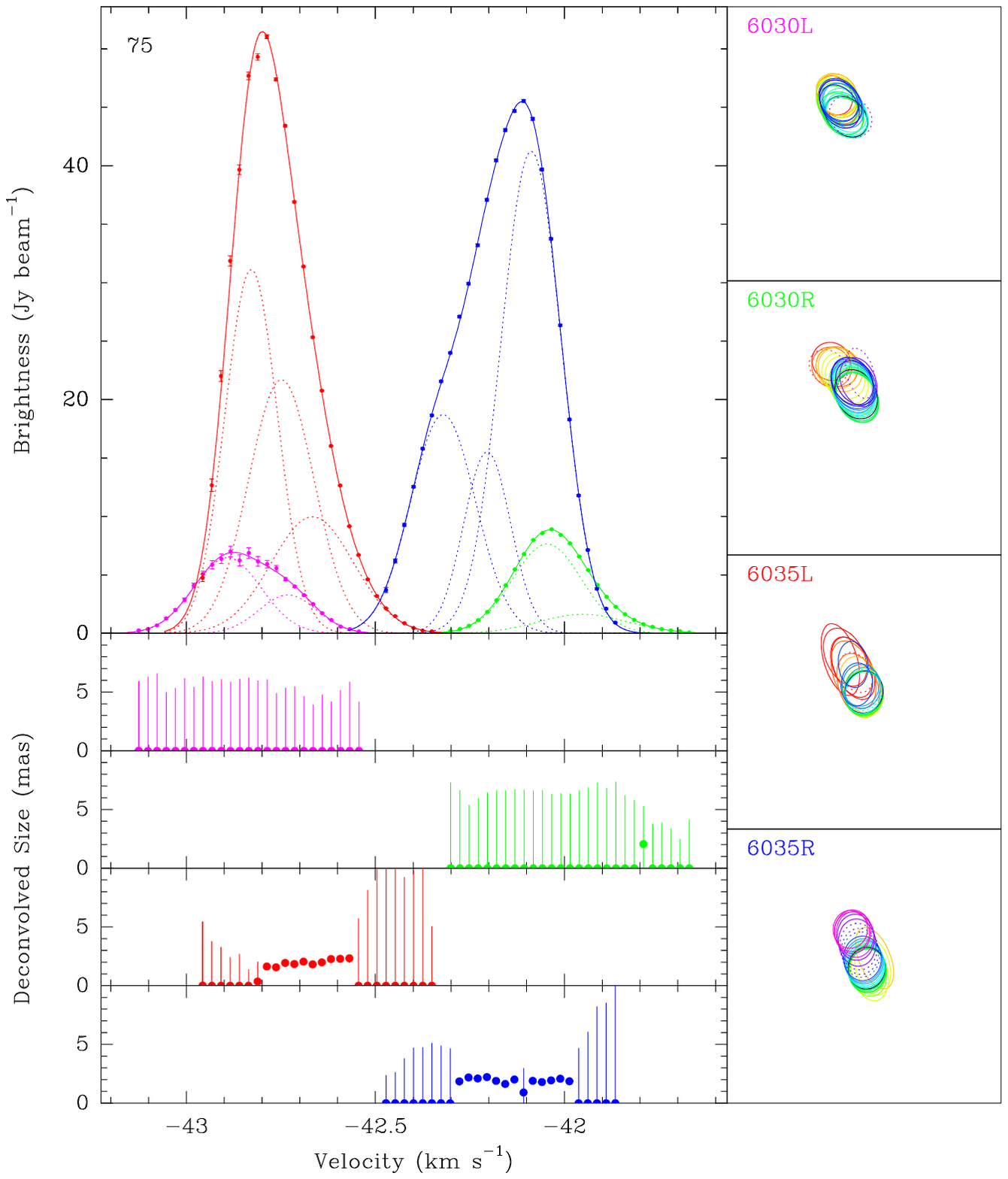}}
\caption{
  Composite plots of selected Zeeman associations.  The upper
  left subpanels show the spectra of 6030 LCP (magenta), 6030 RCP
  (green), 6035 LCP (red), and 6035 RCP (blue) emission (dots with
  error bars), along with individual (dotted lines) and total (solid
  lines) Gaussian fits.  Zeeman association number is shown in the
  upper left.  The bottom subpanels show the deconvolved spot size in
  each velocity channel for each transition/polarization (same
  colors).  The nominal deconvolved spot size is shown as a dot, while
  the range of allowed deconvolved sizes is indicated by a vertical
  line.  The subpanels on the right show the positions of the masers
  in each velocity channel.  Boxes are 20~mas on a side, centered on
  the position of brightest emission in any transition and
  polarization.  Relative positions are correct across
  transitions/polarizations.  Ellipses are one-fifth the size of the
  undeconvolved fitted elliptical Gaussians.  Color indicates velocity
  relative to the channel of peak emission (shown in black).  Dotted
  ellipses represent channels with a signal-to-noise ratio of less
  than 10 or in which emission is blended with a nearby feature (not
  shown).
  }
\label{fig-quads}
\end{figure*}

When the minimum deconvolved spot size is greater than zero, it is
still not clear whether we are detecting the finite size of a single
maser spot or a blend of two unresolved, spatially separated spots.
Figure~\ref{fig-quads} shows deconvolved spot sizes of masers in
several Zeeman associations.  In two of these cases (53 and 75), the
6035~MHz spots appear to have finite size near line center.  However,
the line profiles are not well fit by a single Gaussian component.
Our data cannot distinguish between a single spatially extended maser
with a non-Gaussian spectral profile or a composite of several smaller
masers with Gaussian spectra.  In fact, it is not entirely clear
whether these two cases are different, depending on what ``a'' maser
is (\S \ref{maser}).

\subsection{Saturation}
\label{saturation}

The saturation temperature of a maser is given by
\[
T_s = \frac{h\nu}{2k}\frac{\Gamma}{A} \frac{4\pi}{\Omega},
\]
where $\Gamma$ is the decay rate, $A$ is the Einstein coefficient, and
$\Omega$ is the beaming angle.  Taking typical values of $\Gamma
\approx 0.15$~s$^{-1}$ and $\Omega \approx 10^{-2}$, the saturation
temperature for a 6035~MHz maser is $1.7 \times 10^{10}$~K
(\citealt{reid88}; D98).  The brightest 6035~MHz maser we detect, with
a flux density of 579~Jy in a maximum deconvolved size of 0.7~mas,
corresponds to a brightness temperature of $5.7 \times 10^{13}$~K and
is clearly saturated.  Its line width of 0.2~km\,s$^{-1}$ indicates
that saturated rebroadening to the thermal line width does not occur.
Taking 2~mas as a reasonable upper size limit on a more typical 1~Jy
maser results in a brightness temperature of $1.2 \times 10^{10}$~K.
This is comparable to the highest reasonable saturation temperature
for a filamentary maser beamed toward the observer using a maximum
amplification length of $10^{15}$~cm, the clustering scale of
ground-state masers in W3(OH) as well as other interstellar OH maser
sources \citep{reid80,fish06}.  D98 reach similar conclusions about
the saturation of the 6.0~GHz masers in W3(OH) but note significant
time variability in maser flux densities compared to observations
taken 17 years prior.  We note that long-term maser variability is
common to a wide variety of maser species and transitions, including
some very bright and clearly saturated masers
\citep[e.g.,][]{garay89,alakoz05,fish07}, and therefore does not
constitute prima facie evidence against saturation.  Another piece of
evidence supporting saturation is provided by the preserved flux
ratios between circularly polarized Zeeman components at 6030 and
6035~MHz (Fig.~\ref{fig-zeeman-ratios}), which might not be expected
if the masers were still operating in the unsaturated (i.e.,
exponentially amplifying) regime.

\subsection{Velocity Gradients}
\label{gradients}

Maser velocity gradients were measured following a method similar to
that used by FBS.  In most cases, the gradients were determined
algorithmically.  We identified the nearest spectral channels on
either side of the peak such that the brightness in those channels was
less than half peak brightness.  The velocity gradient was determined
by dividing the LSR velocity difference of the two channels by the
positional difference between them.  Gradient magnitudes and position
angles are reported in Table \ref{tab-spots}.  Note that velocity
gradients are inversely proportional to positional gradients, so a
maser whose position changes rapidly as a function of LSR velocity
will have a small velocity gradient (as measured in
km\,s$^{-1}$\,mas$^{-1}$; 1~km\,s$^{-1}$\,mas$^{-1}$ =
$10^5$~km\,s$^{-1}$\,pc$^{-1}$).

In some cases, spectral channels nearer to the peak channel were used
to compute the gradient instead, for one or more of the following
reasons: sidelobe contamination from a nearby strong maser feature,
spatial blending with another maser feature, spectral blending with
another feature at approximately the same location but at a different
velocity, significant curvature of the locus of channel centroid
positions, and poor determination of channel centroid positions in
instances of low signal-to-noise detections.  The use of spectral
channels close to the peak channel in these instances preserves the
intent of computing the local velocity gradient at line center.
Velocity gradients were determined for a total of 178 masers out of
the 292 detected, 37 of which required using a narrower range of
spectral channels than the range of half-peak brightness.

\begin{figure}[t]
\resizebox{\hsize}{!}{\includegraphics{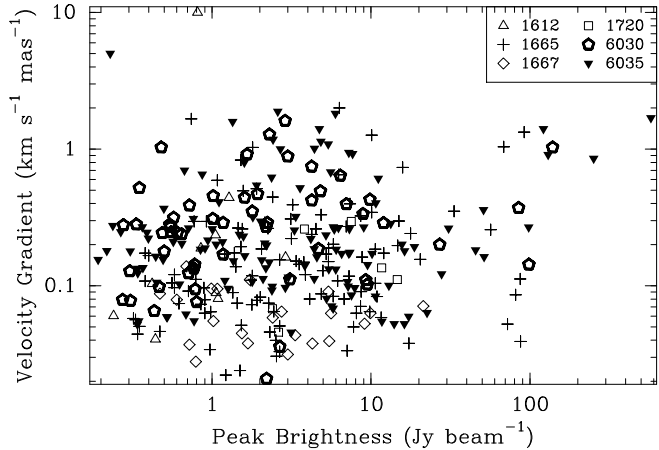}}
\caption{Magnitude of velocity gradient as a function of peak
  brightness.  Bold symbols are from the present work; symbols in
  normal weight are from ground-state data from FBS.  There does
  not appear to be a correlation between the magnitude of the velocity
  gradient and the maser brightness.  Velocity gradients at 1667~MHz
  are systematically smaller than in the other transitions.
\label{fig-grad-flux}
}
\end{figure}

As with the ground-state transitions, there is no correlation between
the brightness of a 6.0~GHz maser and its velocity gradient
(Fig.~\ref{fig-grad-flux}).  The range of magnitudes of velocity
gradients in the 6030 and 6035 MHz transitions is comparable to that
of the ground-state transitions.  However, neither the 6030 nor the
6035 MHz masers display a propensity to occur only in regions of small
velocity gradient (large positional gradient), as do 1667 MHz masers
(FBS).

\begin{figure}[t]
\resizebox{\hsize}{!}{\includegraphics{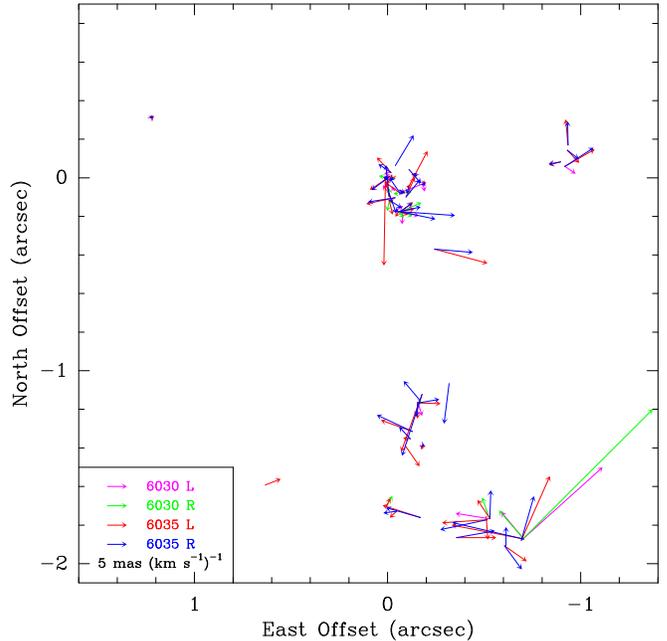}}
\caption{Map of positional gradients of spots with peak brighness
  greater than 1~Jy~beam$^{-1}$.  Arrows point in the direction of
  change with increasing line-of-sight velocity.  Arrow length is
  proportional to the positional gradient, with sample vectors shown
  in the bottom left.  Positional gradients are shown in lieu of
  velocity gradients (their inverse) to illustrate the excellent
  alignment of gradients of members of a Zeeman association (where
  arrow tails are nearly coincident).  Positional gradients are
  typically smaller (velocity gradients are larger) in the cluster
  near the origin.
\label{fig-grad-map}
}
\end{figure}

\begin{figure}[t]
\resizebox{\hsize}{!}{\includegraphics{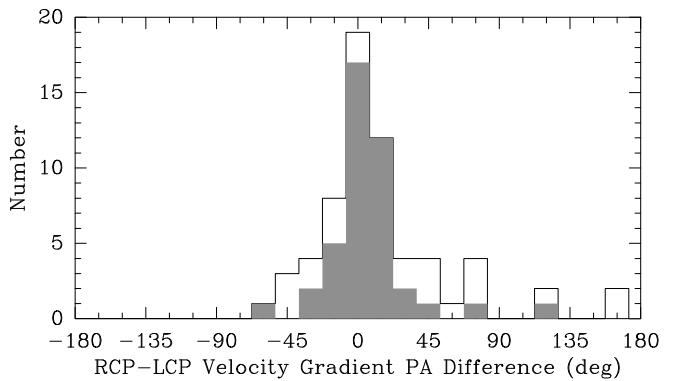}}
\caption{Histogram of position angle differences between the velocity
  gradients of the RCP and LCP components of a Zeeman pair.  The
  distribution is strongly peaked near $0\degr$, indicating that
  components of a Zeeman pair have correlated velocity gradients.  The
  shaded area indicates the histogram for only those Zeeman pairs
  whose average velocity gradient is less than
  0.3~km\,s$^{-1}$\,mas$^{-1}$.  Since these spots have larger
  positional gradients, the error in determining position angles is
  smaller.
\label{fig-grad-histo}
}
\end{figure}

Figure~\ref{fig-grad-map} shows the \emph{positional} gradient of each
maser spot for which it could be determined.  The positional gradient
is the natural observed quantity (the change in centroid position as a
function of LSR velocity, measured in mas\,(km\,s$^{-1}$)$^{-1}$) and
is the inverse of the velocity gradient (measured in
km\,s$^{-1}$\,mas$^{-1}$).  The error in determining the position
angle of the gradient is small for a large change in the maser spot
centroid as a function of LSR velocity (hence our decision to present
positional gradients in preference to velocity gradients, where the
largest gradients have large uncertainties in position angle).  As
seen in the ground-state (FBS), velocity gradients are typically
larger (i.e., the positional gradient is smaller) in the cluster near
the origin.  Positional/velocity gradients are strongly correlated in
a Zeeman association, both within a Zeeman pair
(Fig.~\ref{fig-grad-histo}) and between Zeeman components in the two
different transitions (6030 and 6035~MHz).  This correlation can be
seen visually in the example Zeeman associations in
Figure~\ref{fig-quads}.  Velocity gradients show a stronger
correlation between Zeeman components at 6.0~GHz than in the ground
state (FBS), likely due to the smaller spatial separation between
Zeeman components.  The correlation of velocity gradients is stronger
still if only those pairs with small velocity gradients (i.e., large
positional gradients) are considered, because small errors in
determining the spot center in each channel are less important for
these masers.  However, velocity gradients do not show large-scale
organization (unlike proper motions), suggesting that they trace local
phenomena.

\begin{figure}[t]
\resizebox{\hsize}{!}{\includegraphics{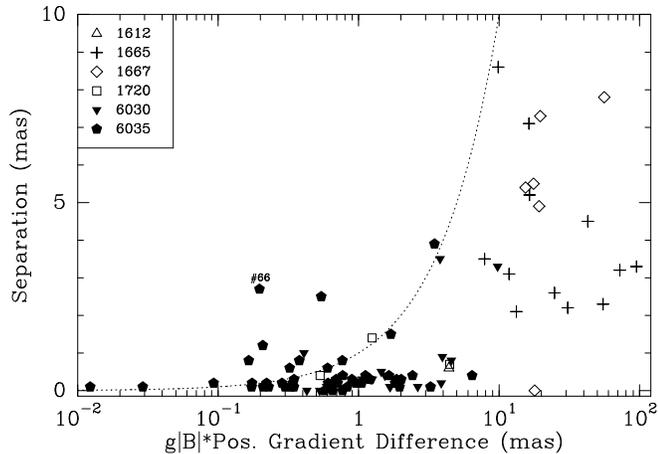}}
\caption{Separation of components of a Zeeman pair plotted against the
  expected spatial separation from the magnetic field.  The abscissa
  is the velocity splitting due to the Zeeman effect multiplied by the
  vectorial difference of the positional gradients of the LCP and RCP
  components of a Zeeman pair.  (Zeeman pairs with separations greater
  than 10~mas are not plotted, since these may be ``Zeeman cousins''
  rather than true Zeeman pairs.)  The dotted line indicates the locus
  of points for which the abscissa and ordinate are equal.  Most spots
  fall under this line, indicating that $g|B|$ times the positional
  gradient difference is in most cases an upper limit to the expected
  spatial separation of Zeeman components.  The largest outlier,
  labelled by Zeeman association number, is discussed in
  \S\ref{gradients}.
\label{fig-sep-gBgrad}
}
\end{figure}

Figure~\ref{fig-sep-gBgrad} shows the transverse separation between
components of a Zeeman pair as a function of the Zeeman splitting
($g|B|$, where $g$ is the Zeeman splitting coefficient in
km\,s$^{-1}$\,mG$^{-1}$ and $B$ is the magnetic field strength in mG)
multiplied by the magnitude of the vectorial difference in positional
gradients in the plane of the sky (mas\,(km\,s$^{-1}$)$^{-1}$).  This
product represents the transverse splitting expected in the presence
of a positional/velocity gradient due to Zeeman splitting.  The LCP
and RCP components of Zeeman pairs usually have a separation
comparable to or smaller than this value.  (The largest outlier in
this plot is the 6035~MHz pair in Zeeman association 66, whose
magnetic field as derived from Zeeman splitting is significantly
different from that of the nearly coincident 6030~MHz pair, suggesting
that physical conditions may change rapidly over a small spatial scale
at this location.)  Thus, the maximum separation between components of
a Zeeman pair is roughly proportional to three quantities: the
magnetic field strength, the Zeeman splitting coefficient, and the
positional gradient (the inverse of the velocity gradient).  This
suggests an explanation for Figure \ref{fig-seps-freq}: the separation
between components of a Zeeman pair increases with the Zeeman
splitting coefficient, but the 1667~MHz Zeeman pairs are even more
widely separated due to the relatively large positional gradients
(i.e., small velocity gradients, as in Fig.~\ref{fig-grad-flux}) seen
in this transition.

\subsection{Linear Polarization}
\label{linear}

Because our data were not correlated to produce cross-polarization (RL
and LR) products, we cannot claim detections of linear polarization.
However, we can comment on the lack of linear polarization from the
signals in the RR and LL correlations alone.  A maser spot with
significant linear polarization (or unpolarized emission) would
produce a detection in both LCP and RCP at the same LSR velocity.  In
the limit of a 100\% linearly polarized spot, as would be expected for
a pure $\pi$-component, the maser fluxes in both circular
polarizations would be equal.

Of the 292 masers we detect, only six pairs of maser spots show this
signature.  Spots 253/256 constitute the clearest detection and may be
due to a high linear polarization fraction in the 6035~MHz LCP
reference feature.  Much weaker linear polarization may be seen in the
coincident 6030~MHz LCP feature (spots 251/254).  In two other cases
(spots 168/170 and 245/246) the peak brightness in the weaker
polarization is less than 0.4\% of that of the brighter polarization,
which could be due to polarization leakage.  Two weaker features at
the same LSR velocity each have more equal LCP/RCP fluxes: spots
193/196 and 282/283.  The former, located in cluster C with large
magnetic field strengths, is likely due to linear polarization.  The
latter is located in cluster SE, which also hosts a 1612~MHz maser
with large linear polarization \citep{wright04b}.  However, we cannot
rule out the alternative interpretation of Zeeman components separated
by an undetectably small magnetic field, since the magnetic field
strengths in cluster SE are smaller than throughout most of the
source.

Thus, linear polarization appears to be rarer in the 6.0~GHz masers in
W3(OH) than in the ground-state masers
\citep{garciabarreto88,wright04a,wright04b}.  We do not note any
6.0~GHz linear polarization candidates in the NW cluster despite the
fact that 1665~MHz masers in this cluster show linear polarization,
although the 6.0~GHz masers are much weaker here than are the
ground-state masers \citep{garciabarreto88}.  We also do not detect
any $\pi$-component candidates.  \citet{wright04a} find several spots
at 1665~MHz with larger linear polarization than circular but also do
not find any convincing $\pi$-component candidates.

\section{Structure of W3(OH)}
\label{structure}

\begin{figure}
\resizebox{\hsize}{!}{\includegraphics{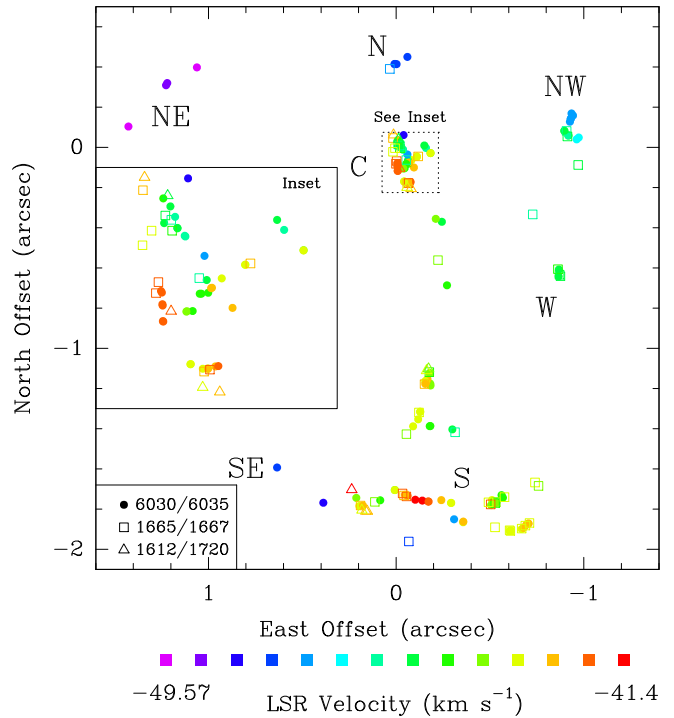}}
\caption{Systemic velocities of Zeeman pairs (i.e., average of LSR
  velocities of RCP and LCP components) in the six masing transitions
  at 1.6 and 6.0~GHz.  Ground-state satellite-line masers (triangles)
  are seen only near the torus feature, while ground-state main-line
  masers (squares) are seen throughout W3(OH), excepting the NE
  cluster.} 
\label{fig-zeeman-vel}
\end{figure}

Figure \ref{fig-zeeman-vel} shows the systemic velocities of Zeeman
pairs in W3(OH).  These velocities are uncontaminated with shifts due
to the Zeeman effect, which can be quite large (several kilometers per
second) in the ground-state transitions.  Several key features stand
out.  First, there is a large range of velocities in the central
cluster, but there is also a clear gradient such that the velocity
decreases toward the northwest.  Second, velocities are clearly
blueshifted in clusters NE, SE, N, and, to a lesser extent, NW.
Third, many of the masers in S are redshifted, and maser velocities
span a large range between S and SE.

Based on these features, the magnetic field
(Fig.~\ref{fig-zeeman-map}), and the distribution of 4765~MHz OH and
6668~MHz methanol emission \citep{harveysmith05,harveysmith06}, we
advance the following model for W3(OH).  There is a ring-shaped
structure, which we shall henceforth refer to as ``the torus''
(Fig.~\ref{fig-map-all}), which is well traced by the 6.0~GHz masers.
The torus may be tracing a circular shock rather than, for instance, a
density disk, but the masers nevertheless exhibit signs of
counterclockwise rotation as viewed from the Earth.  The structures
labelled NW and S trace shocked molecular material in advance of the
ionization front.  The line of masers in S is seen to intersect the
masers in the torus in projection and may be tracing a shock that
intersects the torus in the third dimension and interacts with it.
The masers in cluster N are found at the southern edge of a dense
molecular condensation, while cluster C contains a similar, perhaps
denser, condensation located just in front of the \ion{H}{2} region.
The masers toward the NE are associated with the champagne flow.  The
remainder of this section is devoted to motivating this model.

The inner edge of the torus is delineated by the 6.0~GHz masers, where
conditions (such as temperature or far infrared flux) most favorable
to excited-state emission are found.  The apparent ellipse is roughly
half as wide (east-west) as it is tall (north-south), indicating an
inclination angle of $\sim 30\degr$ to the line of sight, if circular,
although the northern extent of the torus is uncertain.  Much of the
excited-state 4765~MHz emission is found in this region, including an
extended filament oriented along the inner edge of the torus
\citep{harveysmith05}.  This contrasts with the 1665~MHz masers, which
occur predominantly to the west of this edge, with the exception of
the cluster near the origin.  (Some 6035~MHz masers are seen as well
to the west of the inner edge of the torus, but these masers are
weaker than those appearing on the edge itself.)  All 26
satellite-line (1612 and 1720~MHz) masers in W3(OH) detected in FBS
are found near this torus, despite the presence of other regions of
strong main-line emission to the northwest and south.  The
satellite-line masers (open triangles in Figure~\ref{fig-zeeman-vel})
in the south have velocities consistent with being part of structure S
rather than the torus itself, although the interaction between the two
structures may be important.  For completeness, we also note that
\citet{wright04b} detect a weak 1720~MHz Zeeman pair in the cluster at
the western edge of the bright continuum emission and a weak highly
redshifted ($-41.48$~km\,s$^{-1}$) 1612~MHz maser in the southern line
of maser spots to the west of the torus.

The torus appears to be rotating in the counterclockwise sense.  The
masers projected along the face of the UC\ion{H}{2} region have LSR
velocities of $\sim -44$~km\,s$^{-1}$, slightly blueshifted from the
presumed systemic velocity of $-46$ to $-45$~km\,s$^{-1}$
\citep{welch87,keto95}, a finding that is true for the 6668, 12178,
and 23121~MHz methanol masers as well
\citep{menten88,moscadelli99,harveysmith06}.  The masers in cluster SE
are very blueshifted, which would be consistent if their location is
on the back side of the star forming region, since counterclockwise
rotation would give these masers a net velocity toward the observer
compared to systemic.  Cluster C is more complicated due to the
gradient in LSR velocity to the northwest, a phenomenon noted as well
in methanol masers at 6668 and 12178~MHz
\citep{moscadelli99,harveysmith06}, which \citet{moscadelli02} model
as conical expansion.  It is not a priori clear whether the torus edge
would be expected to pass through the center of this cluster or its
inner (eastern) edge.  We note, however, that all satellite-line (1612
and 1720~MHz) masers in cluster C appear along or near the inner edge,
which is redshifted with respect to the systemic velocity of W3(OH)
(see inset in Fig.~\ref{fig-zeeman-vel}).  The sum of these properties
is consistent with counterclockwise rotation, which is opposite to the
direction of rotation assumed by \citet{wright04a} from 1665~MHz
observations.  However their interpretation is based on 1665~MHz
demagnetized velocities of $\approx -43$~km\,s$^{-1}$ to the south,
which we argue are associated with feature S (not in the torus) rather
than feature SE (part of the back side of the torus).  Ultimate
confirmation of the rotation will depend on a further epoch of proper
motions, as Figure~\ref{fig-motions} does not provide information on
the motions of the newly-detected 6.0~GHz masers in the southeast of
W3(OH).

Based on maser velocities, \citet{wright04b} speculate that the
magnetic field is oriented toward the observer in the southeast, a
finding we confirm from direct examination of 6.0~GHz Zeeman pairs.
They also argue that these masers are located on the back side of the
\ion{H}{2} region, again based on their LSR velocities, which are
blueshifted by several kilometers per second compared to the nearest
masers as seen in projection (Fig.~\ref{fig-map-all}).  While the
\ion{H}{2} region is probably optically thick at $\lambda = 5$~cm
\citep[e.g.,][]{baudry93}, these masers are located south of or
coincident with the \ion{H}{2} region limb, so the ionized material is
plausibly not sufficiently optically thick to prevent detection of
masers from behind here, although the optical depth of the \ion{H}{2}
region may partially explain the weakness of these masers.  The
magnetic field polarity, negative behind the \ion{H}{2} region and
positive in front of it, indicates that the toroidal component of the
magnetic field is oriented in the counterclockwise sense.

The torus model also can explain the filamentary 4765~MHz emission
seen by \citet{harveysmith05}.  Diffuse, low-gain maser emission is
seen along the torus, with peak emission near and north from
declination offset $-1\arcsec$ at $-44$ to $-43$~km\,s$^{-1}$.
(Emission is also seen on the eastern edges of clusters C and W.)  The
velocity is consistent with nearby OH and methanol masers, and the
location is near where the rotating torus would be expected to have no
net line-of-sight velocity with respect to the Earth.  The presence of
the diffuse maser emission may therefore be an indicator of a region
of line-of-sight velocity coherence, a probable prerequisite for
significant low-gain amplification.

We note that the torus does not necessarily trace a circumstellar
disk.  Linear or arc-like structures with regular velocity structures
suggestive of rotation have been noted in other maser sources,
especially in methanol, and interpreted as circumstellar disks
\citep{norris93,norris98}.  However, mid infrared imaging consistently
indicates that these putative disk sources are associated with
outflows and/or shocks instead \citep{debuizer03}.  It may be possible
to obtain these maser structures if a shock oriented mostly
perpendicular to the line of sight propagates through a rotating cloud
\citep{dodson04}.  The torus may therefore be tracing the interaction
of an expanding shock with a rotating molecular structure, with a
rapid fall-off of density to the east.  The distribution of ammonia
absorption, seen over the western half of the \ion{H}{2} region in
W3(OH) but not the eastern half, supports this interpretation
\citep{guilloteau83,reid87}.  This pattern of absorption is seen in
excited-state OH and formaldehyde as well
\citep{guilloteau85,dickel87}.

The inclination of the torus indicates that the western side of the
torus is in front of the \ion{H}{2} region, or equivalently that the
eastern side of the \ion{H}{2} region is closer to the observer than
the western side).  The density of the surrounding material is lower
to the east than to the west, as confirmed by several pieces of
evidence.  First, the density of ammonia seen in absorption falls off
from west to east across the face of W3(OH), with particularly low
values obtained toward the southeastern (SE) masers \citep{reid87}.
Second, the ground-state masers in groups NW and S appear to be
oriented along the limbs of the \ion{H}{2} region where the radio
continuum emission gradient is steepest, suggesting compression due to
expansion into higher-density material.  Third, the expansion of the
\ion{H}{2} region is fastest toward the east \citep{kawamura98}.
Fourth, the champagne flow on the eastern side necessarily indicates a
lower density to the east.  This may also explain why the northeastern
(NE) masers are especially blueshifted -- they are located closest to
the champagne flow \citep{wilner99}, where the ambient density is
lowest, so they are accelerated most by the expansion of W3(OH).
Indeed, an accelerated champagne flow model is supported by the
recombination line study of \citet{sams96}.  Fifth, the magnetic field
along the putative maser torus is fairly consistently about 5~mG
across the face of the \ion{H}{2} region, except in cluster C where it
is larger.  This would be expected if the density here is higher,
which it almost certainly must be in order to support both very strong
6.0~GHz maser emission as well as 13.4~GHz masers
\citep{baudry98,cragg02,wright04a}.  But to the south along this arc,
the magnetic field along the torus drops, below 5~mG for all six
positive field values at 6.0~GHz (Zeeman associations 78 and 80--83).
The two Zeeman pairs (numbers 84 and 85) indicating a negative
magnetic field value both give magnetic field values of approximately
$-1$~mG, indicating that the magnetic field (and therefore density)
drop substantially to the southeast.  Strong magnetic fields at
1612~MHz at the southern tip of the torus (Zeeman pairs 52--54 in FBS)
complicate the picture somewhat, but it is probable that these masers
are actually associated with the structure labelled ``S'' in
Figure~\ref{fig-map-all}, given that these masers are redshifted
compared to the nearby masers in the torus and especially compared to
the SE masers.

Cluster C is clearly a very energetic region.  In addition to the OH
masers at 1.6, 4.7, 6.0, and 13.4~GHz, there are weaker, extremely
rare masers in the highly-excited 7820 and 8189~MHz OH transitions
\citep{baudry93}.  This is additionally the region with the brightest
methanol maser emission \citep{harveysmith06}.  It is traditionally
assumed that the main excitation source in the ionized region is near
cluster C, based on the maser properties, higher magnetic field, and
23~GHz continuum peak \citep{guilloteau83}.  It is perhaps not
coincidental that the region of largest magnetic field strength occurs
in cluster C, not far from the northern pinch.

Cluster N is likely at the periphery of a very dense region.  It is
probable that a region of high density is responsible for the
prominent pinch in the \ion{H}{2} region (see Fig.~\ref{fig-map-all}).
The 6668~MHz methanol emission also provides indirect evidence of the
high density.  Bright 6668~MHz maser emission is cospatial with OH
maser emission everywhere in W3(OH) \emph{except} at cluster N
\citep{harveysmith06}.  The methanol maser emission nestles well into
the pinch, and the OH masers (1665, 6030, and 6035~MHz) are found
between the \ion{H}{2} region and the dense clump.  This scenario is
supported by the model of \citet{cragg02}, who find that warm material
can support 6668~MHz methanol masers at densities higher by a factor
of several than 1665~MHz masers, while other ground-state transitions,
not seen in cluster N, require still smaller densities for maser
activity.  The 6.0~GHz masers in the region may sample slightly
cooler, though still fairly dense, material located on the inside of
the higher-density methanol clump.

Besides the maser torus, large quantities of masers are seen in two
regions: northwest (NW) and south (S) of the ionized region.  In both
of these regions, the limb of the ionized emission is very sharply
defined (i.e., the contours of the radio continuum of the \ion{H}{2}
region are packed closely together in Fig.~\ref{fig-map-all}), and the
distribution of maser spots is preferentially along, but offset about
400~AU from, the edge of the ionized region.  This is consistent with
expectations for the expansion of a $D$-type shock \citep{shu92}.
When the expansion speed of the \ion{H}{2} region is less than twice
the sound speed of the ionized region, both an ionization front and a
shock front will be produced.  The shocked neutral material between
the fronts will often have conditions favorable to maser production
\citep{elitzur78}.  The typical expansion speed of both the
ground-state masers \citep{bloemhof92,wright04a} and the \ion{H}{2}
region itself \citep{kawamura98} is 3--5~km\,s$^{-1}$, substantially
smaller than the sound speed in the ionized gas \citep[$\sim
10$~km\,s$^{-1}$,][]{yorke86}.  It is noteworthy that the southern
line of masers and the northwestern clump are offset directly from the
periphery of the \ion{H}{2} region in areas where no expansion is seen
in the \citet{kawamura98} data (see their Fig.~3).  The morphology and
velocity structure of the masers in S are complicated and may be
indicative of multiple shocks (or possibly a single shock with several
regions of maser activity seen in projection) or energy input from an
unseen source.

The masers indicating a negative magnetic field to the northeast are
seen against continuum emission, albeit weaker emission than seen over
most of the face of W3(OH).  It is not a priori clear whether these
masers are located in front of or behind the continuum emission.
While these masers are the most blueshifted of any 6.0~GHz masers seen
in W3(OH), the lone Zeeman pair indicating a positive magnetic field
in the region (Zeeman association 86, see Fig.~\ref{fig-minus-B}) is
intermediate in LSR velocity to the triad of Zeeman pairs indicating
negative magnetic fields, suggesting that each of the four Zeeman
pairs originate from the same side (front or back) of the continuum
emission as the others.  It is possible that the magnetic field wraps
around the front face of the ionized region here \citep[as in Fig.~7
of][]{bourke01}, resulting in a reversal of the line-of-sight magnetic
field direction.  The density of the material in the NE cluster is
clearly lower, as is supported by both the submilligauss magnetic
field strengths and the existence of the champagne flow.

\section{Multi-transition Maser Comparison}
\label{multitrans}

Many theoretical models conclude that 6030 and 6035~MHz masers are
excited under similar conditions
\citep[e.g.,][]{gray91,pavlakis00,cragg02}.  Indeed, we find very
close spatial association of 6030 and 6035~MHz masers, with typical
separations being comparable to the separation within a
single-frequency Zeeman pair and typical velocity differences much
less than a line width after correcting for the Zeeman effect (\S
\ref{zeeman}).  Masers at these two frequencies in Zeeman associations
usually indicate similar magnetic field strengths and velocity
gradients.  The remarkable match in properties between 6030 and
6035~MHz masers admits the stronger conclusion that excitation of 6030
and 6035~MHz masers proceeds under similar conditions.  It appears
that, in fact, in many cases 6030 and 6035~MHz masers trace the same
volume of gas, a conclusion that is not true for ground-state masers.

We do not find that 6.0~GHz masers prefer regions of either large or
small velocity gradients, unlike 1667~MHz which exist almost
exclusively in regions of small velocity gradients (FBS).  This stands
in contrast to the prediction of \citet{pavlakis00} that inversion of
both 6.0~GHz main lines requires a small velocity gradient.  They
model masers as cylinders of diameter $10^{15}$~cm with linear
velocity gradients both along and across the cylinder (their $V =
1$~km\,s$^{-1}$ corresponds to a velocity gradient of
3000~km\,s$^{-1}$~pc$^{-1}$ or 0.03~km\,s$^{-1}$\,mas$^{-1}$ at the
distance of W3(OH)).  It is surely the case that the transverse size
of the 6.0~GHz masers in W3(OH) is substantially smaller if they are
cylindrical.  The deconvolved spot sizes of our masers are consistent
with being pointlike, indicating that actual maser sizes are at least
a factor of several times smaller than the 7~mas ($2 \times
10^{14}$~cm) synthesized beam size.  What is unclear, however, is
whether velocity gradients in the plane of the sky correlate with
velocity gradients along the line of sight.  An observational bias may
exist because a large velocity gradient along the line of sight would
decrease the path length of velocity coherence for an OH maser,
potentially rendering it undetectable; in principle, no such bias
exists for velocity gradients transverse to the line of sight.  On the
other hand, this effect may be offset by velocity redistribution
\citep[][see also \S \ref{maser}]{field94}.

Models also predict that 6030~MHz masers should be slightly weaker
than 6035~MHz masers, a finding strongly confirmed in Zeeman
associations in W3(OH) (Fig.~\ref{fig-zeeman-ratios}) as well as
surveys of other sources \citep[e.g.,][]{caswell03}.  Effectively all
6030~MHz masers are accompanied with nearby, stronger 6035~MHz maser
emission.  The largest spatial separation between a 6030~MHz maser and
the nearest 6035~MHz maser is less than 40~mas, but the 6030~MHz
masers (Zeeman association 10 and spot 27) are especially weak.  The
6030~MHz masers in Zeeman association 81 are also relatively weak ($<
1$~Jy\,beam$^{-1}$) and separated from the nearest 6035~MHz by almost
20~mas.  In all of these cases where 6030~MHz masers are found but
6035~MHz masers are absent, there is nearby 1665~MHz maser emission,
matching a prediction by \citet{pavlakis00}.  Three of these five
6030~MHz masers are not so weak as to prevent an estimate of the
velocity gradient, which ranges from 0.027 to
0.054~km\,s$^{-1}$\,mas$^{-1}$.  Caution is warranted in interpreting
apparent velocity gradients of weak masers (the error in determining
the position centroid in each channel is inversely proportional to the
signal-to-noise ratio), but the velocity gradients are nevertheless
small, although at the low end of the range predicted by
\citet{pavlakis00}.

Close spatial comparison with the ground-state masers is difficult
because the absolute position of the data in FBS is not known, and the
ground-state and 6.0~GHz maps cannot be aligned to milliarcsecond
accuracy based on the location of Zeeman pairs.  Nevertheless, the
overall distribution of the masers match closely enough (within a few
tens of milliarcseconds) to permit some general statements to be made.
The 1720~MHz masers in W3(OH) do appear to be located near (within
10--20~mas of) 6035~MHz masers, as predicted
\citep[e.g.,][]{gray91,cragg02}.  Zeeman pair 42 of FBS (6.6~mG)
matches very well with Zeeman association 70 of the present work
(6.5~mG), and spot 266 of the present work would indicate a 7.6~mG
magnetic field if shifted from the systemic velocity of Zeeman pair 47
of FBS (6.8~mG).  The other satellite-line masers at 1612~MHz also
appear only in the general vicinity of 6035~MHz masers, although the
spatial separation from the nearest 6035~MHz masers is at times larger
than is true for the 1720~MHz masers.  Masers in the 1667~MHz
transition align with 6.0~GHz masers primarily in the southwest of
W3(OH).  Where 1667~MHz masers overlap with 6.0~GHz masers, they do so
preferentially with associations in which both 6030 and 6035~MHz
masers congregate.  Nevertheless, the presence of both 6.0~GHz main
line transitions together does not predict 1667~MHz masers; for
instance, numerous 6030/6035~MHz associations exist in the bright
cluster near the origin and to a lesser extent in the center and
northwest of W3(OH), but no 1667~MHz emission is found in these
regions.  In comparison, 1665~MHz masers are ubiquitous and appear in
large quantities both in regions lacking 6.0~GHz emission as well as
in regions where 6.0~GHz masers are plentiful, either as 6035~MHz
masers alone or as 6030/6035~MHz associations.

All satellite-line masers detected by FBS are seen in the torus.  This
is likely due to the fact that the torus traces the inner edge of a
region of high density.  The differences between the locations of the
1612 and 1720~MHz masers may be related to the necessary pump
processes.  The 1720~MHz masers are found in two regions: cluster C
and the edge of the torus near a Declination offset of $-1100$~mas.
\citet{cragg02} find that 1720~MHz emission requires a low kinetic
temperature (or high dust temperature) and high OH specific column
density, equal to the volume density of OH divided by the velocity
gradient along the amplification length.  This condition is likely
achieved in cluster C due to a high molecular column density.  Farther
south, the high specific column density is likely partially due to
velocity coherence along the line of sight, given the peak in 4765~MHz
OH maser emission at this location \citep{harveysmith05}, as discussed
in \S \ref{structure}.  All 1720~MHz masers are found projected atop
the \ion{H}{2} emission, where far infrared photon likely enhances
the dust temperature.  In contrast, \citet{cragg02} find that 1612~MHz
masers also prefer regions of high specific column density but require
a high gas temperature (or low dust temperature).  Masers in the
1612~MHz transition are found in cluster C and where the torus
intersects structure S.  Again, the total column density in cluster C
is very high, although it should be noted that the 1612 and 1720~MHz
masers come from different parts of the cluster, satisfying the
predicted conjugate behavior \citep{cragg02}.  However, the 1612 MHz
masers to the south are all located off the limb of the continuum
emission and (with one possible exception in the SE cluster) are
probably associated with the interaction between the torus and
structure S, which we argue is likely shock-driven and therefore
consistent with a high gas temperature and a low dust temperature.
Thus, the locations of the 1612 and 1720~MHz masers support findings
by \citet{pavlakis96a,pavlakis96b} that at high densities ($n_{H_2} >
10^{6}$~cm$^{-3}$) the 1612~MHz line can be excited by collisions
alone, while inversion of the 1720~MHz line may be supplemented by
infrared pumping.

\section{Maser Substructure and Superstructure}
\label{maser}

It is clear that the masers in W3(OH) have structure on scales smaller
than observed (e.g., \S \ref{deconvolved}).  Other maser transitions
that have been observed in W3(OH), with higher angular resolution
being achieved at higher frequencies, support this conclusion.  Some
1665~MHz masers in W3(OH) are partially resolved at VLBI resolution,
although others show deconvolved spot sizes of 3~mas or less
\citep{reid80,garciabarreto88}.  The 1720~MHz masers show evidence for
sizes less than 1.2~mas \citep{masheder94}.  In other sources, the
smallest intrinsic scales of structure are less than 2~AU (1~mas at
the distance of W3(OH)) \citep{hansen93,slysh01}.  The highly-excited
13441~MHz masers in W3(OH) are as small as 400~$\mu$as
\citep{baudry98}.  Even some 12.2~GHz methanol masers in W3(OH) show
evidence for submilliarcsecond structure \citep{moscadelli02}.

On the other hand, it is also clear that masers in W3(OH) have
larger-scale structure as well.  Most of the 4765~MHz masers are
larger than the synthesized VLBA beam (3--4~mas) and show flux on
larger spatial scales than observable with the VLBA \citep{palmer03}.
The 13441~MHz masers appear in filamentary or arc-like structures up
to several tens of milliarcseconds in size \citep{baudry98}.  Extended
structure is seen in 6.7~GHz methanol masers \citep{harveysmith06} and
12.2~GHz methanol masers as well \citep{moscadelli02}, with smaller
ground-state and 6.0~GHz masers distributed along the filaments
(\citealt{wright04a,wright04b,etoka05}; FBS; this work).

Thus, the structures we see as masers likely both contain substructure
and are themselves the bright, compact peaks of a larger structure.
At our spatial and spectral resolution, the simplistic model of a
maser profile as a narrow Gaussian begins to break down.  Multiple
components, possibly even a continuum of components whose parameters
vary spatially, are required to model a maser line profile (see \S
\ref{profiles} and Fig.~\ref{fig-quads} as well as
\citealt{moscadelli03}).  The question of what exactly constitutes
``a'' maser is not merely a philosophical one, but one that must be
confronted in order to interpret modern observations.  A detailed
understanding of these effects is especially important for observers
attempting to obtain geometric parallax distances from maser motions
with respect to a distant background.  Small changes in maser
substructure may manifest themselves in apparent shifts of the maser
centroid, limiting the accuracy to which such distances can be
measured.

Maser observations with both very high spectral resolution and angular
resolution provide a unique probe of conditions on AU scales.  The
velocity gradients that are seen may represent AU-scale turbulent
fluctuations in the transverse velocity field.  An analysis of the
distribution and line-of-sight velocity scatter of water masers
indicates that the inner scale of supersonic turbulent dissipation
occurs at approximately 1~AU \citep{strelnitski02}, perhaps not
coincidentally the size scale appropriate for OH masers as well.  A
similar analysis of maser transverse velocity gradients may permit
estimation of parameters related to the inner turbulent scale.
Understanding turbulence is important to understanding masers.  Simple
models of turbulence in a homogeneous medium can reproduce many
observed maser properties \citep[e.g.,][]{boger03}, while a real star
forming region contains both turbulence and significant
inhomogeneities in density, temperature, far infrared photon flux, and
other physical parameters.  Future maser observations with very high
spatial resolution, spectral resolution, and sensitivity may
contribute greatly to our knowledge of maser processes.

\section{Conclusions}

1.\ The locations and velocities of the 6.0~GHz masers across the face
and to the southeast of the the continuum emission in W3(OH) support
the model of a dense molecular torus put forth by \citet{guilloteau83}
and \citet{dickel87}.

2.\ There is a reversal of the line-of-sight direction of the magnetic
field in W3(OH), with positive magnetic fields (i.e., oriented away
from the observer) in the west and negative magnetic fields in the
east.  The polarity of the magnetic field and demagnetized OH maser
velocities at the inner edge of the torus suggest that the torus is
undergoing net counterclockwise rotation.

3.\ Masers in the 6030 and 6035~MHz sample the same material.  They
are often found in direct spatial overlap and have nearly identical
demagnetized LSR velocities, magnetic fields, and velocity gradients.
Masers in the 6030~MHz transition are usually accompanied by stronger
6035~MHz emission, indicating that conditions conducive to masing in
the 6030~MHz line are also conducive to masing at 6035~MHz, although
with larger gain in the latter.  Remarkably, even the ratio of
circularly polarized components in a Zeeman pair in one transition is
approximately preserved in the other.

4.\ The intrinsic sizes of 6.0~GHz masers in W3(OH) are considerably
smaller than the synthesized beam size (7 mas = 14 AU).  Line shapes
are in many cases nearly Gaussian with a line width of 0.15 --
0.40~km\,s$^{-1}$, although adequate line shape fitting requires
multiple (possibly a continuum of) Gaussian components.  Most, if not
all, detected 6.0~GHz masers are at least partially saturated.

5.\ Proper motions of 6.0~GHz masers are faster than motions at 1665
MHz \citep{bloemhof92} and appear to be dominated more by rotation
than expansion, a conclusion obtained at 1665 MHz by
\citet{wright04a}.  A clearer picture of the kinematics of the 6.0~GHz
masers will require another epoch of observations with the EVN.  In
particular, tracking motions of the S and SE masers will greatly aid
understanding of the nature and rotation (and possible expansion) of
the torus, while tracking motions of the NE masers will clarify their
relation to the champagne flow.

6. W3(OH) hosts numerous OH masers with a range of densities, magnetic
field strengths, pumping conditions, and gas motions all located
within several hundred AU of the \ion{H}{2} region, complicating
simple interpretation in terms of the origin, structure, and evolution
of W3(OH).

7. Masers are not just spots but trace a continuum of different
structures of larger-scale gas.  Depending on physical conditions,
they may appear most prominently as weak diffuse emission or strong
compact objects.  Future observations with high sensitivity, spectral
resolution, and spatial resolution may allow estimation of turbulent
parameters.

\acknowledgments

The European VLBI Network is a joint facility of European, Chinese,
South African and other radio astronomy institutes funded by their
national research councils.  The National Radio Astronomy Observatory
is a facility of the National Science Foundation operated under
cooperative agreement by Associated Universities, Inc.  We thank
W.~F.\ Brisken for helpful comments in manuscript preparation.


{\it Facilities: EVN}

\clearpage

\LongTables
\begin{deluxetable}{lccrrrrrrc}
\tabletypesize{\tiny}
\tablecaption{Detected 6030 and 6035 MHz Masers in W3(OH)\label{tab-spots}}
\tablehead{
  \colhead{Spot} &
  \colhead{Freq.} &
  \colhead{} &
  \colhead{R.A.~Offset\tnm{a}} &
  \colhead{Decl.~Offset\tnm{a}} &
  \colhead{Velocity\tnm{b}} &
  \colhead{Brightness\tnm{c}} &
  \colhead{Vel.~Gradient} &
  \colhead{P.A.\tnm{d}} &
  \colhead{Zeeman} \\
  \colhead{Number} &
  \colhead{(MHz)} &
  \colhead{Polar.} &
  \colhead{(mas)} &
  \colhead{(mas)} &
  \colhead{(km\,s$^{-1}$)} &
  \colhead{(Jy\,beam$^{-1}$)} &
  \colhead{(km\,s$^{-1}$\,mas$^{-1}$)} &
  \colhead{(deg)} &
  \colhead{Group\tnm{e}}
}
\startdata
  1& 6035 & L & $-$973.59 &      50.73 & $-$46.40 &   0.14 &  \nogradient   & 1 \\
  2& 6035 & R & $-$972.09 &      48.25 & $-$46.13 &   0.19 &  \nogradient   & 1 \\
  3& 6035 & L & $-$961.91 &      40.38 & $-$46.33 &   0.31 &  0.179 & $-$73 & 2 \\
  4& 6035 & R & $-$961.57 &      40.33 & $-$46.04 &   0.62 &  0.103 & $-$73 & 2 \\
  5& 6035 & R & $-$955.93 &      91.60 & $-$46.23 &   0.23 &  \nogradient   & \\
  6& 6035 & R & $-$942.41 &     158.14 & $-$46.50 &   0.30 &  \nogradient   & 3 \\
  7& 6035 & L & $-$942.23 &     159.55 & $-$46.81 &   0.14 &  \nogradient   & 3 \\
  8& 6035 & R & $-$938.62 &     147.91 & $-$46.04 &   0.22 &  \nogradient   & \\
  9& 6035 & R & $-$934.46 &     169.20 & $-$46.84 &   4.57 &  0.162 &     1 & 4 \\
 10& 6035 & L & $-$934.31 &     169.29 & $-$47.15 &   2.29 &  0.149 &     5 & 4 \\
 11& 6035 & R & $-$933.04 &     119.79 & $-$46.50 &   0.21 &  \nogradient   & \\
 12& 6035 & R & $-$930.63 &     143.29 & $-$47.01 &   6.25 &  0.266 &$-$128 & 5 \\
 13& 6035 & L & $-$930.53 &     143.43 & $-$47.35 &   4.19 &  0.221 &$-$137 & 5 \\
 14& 6030 & L & $-$930.46 &     142.57 & $-$47.35 &   0.35 &  0.521 &    35 & 5 \\
 15& 6030 & R & $-$930.43 &     142.64 & $-$46.86 &   0.54 &  0.281 &$-$167 & 5 \\
 16& 6035 & R & $-$926.22 &     128.65 & $-$46.57 &   0.54 &  \nogradient   & 6 \\
 17& 6035 & L & $-$925.97 &     127.52 & $-$46.86 &   0.35 &  0.055 &   169 & 6 \\
 18& 6035 & R & $-$918.33 &      59.01 & $-$45.26 &   2.10 &  0.113 & $-$57 & 7 \\
 19& 6035 & L & $-$918.21 &      59.10 & $-$45.63 &   4.39 &  0.112 & $-$60 & 7 \\
 20& 6030 & L & $-$916.72 &      60.39 & $-$45.89 &   1.18 &  0.289 &$-$124 & 7 \\
 21& 6030 & R & $-$916.68 &      60.30 & $-$45.41 &   0.60 &  0.244 &$-$140 & 7 \\
 22& 6030 & R & $-$896.59 &      82.36 & $-$45.24 &   0.48 &  1.031 &   161 & 8 \\
 23& 6030 & L & $-$896.42 &      82.22 & $-$45.70 &   0.64 &  0.239 &   111 & 8 \\
 24& 6035 & L & $-$896.41 &      82.54 & $-$45.55 &   3.36 &  0.321 &   105 & 8 \\
 25& 6035 & R & $-$896.36 &      82.44 & $-$45.21 &   1.81 &  0.318 &   100 & 8 \\
 26& 6035 & L & $-$889.81 &      97.33 & $-$45.29 &   0.65 &  0.203 & $-$71 & \\
 27& 6030 & R & $-$870.34 &  $-$634.68 & $-$44.97 &   0.13 &  \nogradient   & \\
 28& 6035 & R & $-$867.21 &  $-$607.06 & $-$44.87 &   0.76 &  0.310 & $-$83 & 9 \\
 29& 6035 & L & $-$867.18 &  $-$607.09 & $-$45.04 &   0.94 &  0.216 &$-$151 & 9 \\
 30& 6030 & R & $-$866.91 &  $-$639.65 & $-$44.92 &   0.16 &  \nogradient   & 10\\
 31& 6030 & L & $-$864.76 &  $-$643.47 & $-$45.09 &   0.09 &  \nogradient   & 10\\
 32& 6035 & L & $-$843.16 &  $-$688.73 & $-$45.16 &   0.26 &  \nogradient   & \\
 33& 6030 & R & $-$712.85 & $-$1868.75 & $-$43.03 &   0.35 &  \nogradient   & \\
 34& 6035 & R & $-$705.61 & $-$1872.25 & $-$42.96 &  13.98 &  0.053 &    77 & 11\\
 35& 6030 & R & $-$705.20 & $-$1871.08 & $-$42.83 &   9.48 &  0.102 &    41 & 11\\
 36& 6030 & L & $-$704.96 & $-$1871.24 & $-$43.32 &   9.27 &  0.110 &    40 & 11\\
 37& 6035 & L & $-$704.48 & $-$1872.05 & $-$43.27 &  16.22 &  0.053 &    79 & 11\\
 38& 6030 & L & $-$699.15 & $-$1873.12 & $-$43.27 &   2.67 &  0.036 & $-$48 & 12\\
 39& 6030 & R & $-$696.92 & $-$1875.50 & $-$42.79 &   2.20 &  0.021 & $-$45 & 12\\
 40& 6035 & L & $-$692.53 & $-$1879.25 & $-$43.30 &  11.70 &  0.055 & $-$24 & 12\\
 41& 6035 & R & $-$692.27 & $-$1879.54 & $-$42.93 &  10.67 &  0.084 & $-$16 & 12\\
 42& 6030 & R & $-$679.69 & $-$1892.03 & $-$42.71 &   0.57 &  \nogradient   & 13\\
 43& 6030 & L & $-$678.48 & $-$1893.00 & $-$43.22 &   0.57 &  \nogradient   & 13\\
 44& 6035 & L & $-$669.12 & $-$1898.58 & $-$43.42 &   2.61 &  \nogradient   & 14\\
 45& 6030 & R & $-$668.92 & $-$1898.62 & $-$42.98 &   0.36 &  \nogradient   & (14)\\
 46& 6035 & R & $-$668.77 & $-$1899.20 & $-$43.03 &   2.44 &  \nogradient   & 14\\
 47& 6035 & R & $-$612.52 & $-$1907.52 & $-$43.30 &   2.15 &  0.204 &  $-$1 & 15\\
 48& 6035 & L & $-$611.92 & $-$1907.19 & $-$43.64 &   3.02 &  \nogradient   & 15\\
 49& 6030 & L & $-$606.57 & $-$1907.87 & $-$43.76 &   0.34 &  \nogradient   & 16\\
 50& 6030 & R & $-$606.41 & $-$1907.87 & $-$43.25 &   0.25 &  \nogradient   & 16\\
 51& 6035 & R & $-$605.90 & $-$1907.35 & $-$43.32 &   6.01 &  0.132 &$-$144 & 16\\
 52& 6035 & L & $-$605.69 & $-$1907.22 & $-$43.71 &   9.52 &  0.142 &$-$125 & 16\\
 53& 6035 & L & $-$569.11 & $-$1742.31 & $-$44.95 &   0.27 &  \nogradient   & 17\\
 54& 6035 & R & $-$568.61 & $-$1743.26 & $-$44.68 &   0.21 &  \nogradient   & 17\\
 55& 6035 & R & $-$561.82 & $-$1730.11 & $-$44.68 &   0.26 &  \nogradient   & 18\\
 56& 6035 & L & $-$561.77 & $-$1729.86 & $-$44.92 &   0.52 &  0.059 &    98 & 18\\
 57& 6030 & L & $-$539.12 & $-$1763.73 & $-$44.70 &   0.56 &  \nogradient   & \\
 58& 6035 & R & $-$535.40 & $-$1764.79 & $-$44.32 &   0.95 &  \nogradient   & 19\\
 59& 6035 & L & $-$533.25 & $-$1765.57 & $-$44.53 &   3.96 &  \nogradient   & 19\\
 60& 6030 & R & $-$531.76 & $-$1767.49 & $-$44.19 &   0.31 &  \nogradient   & \\
 61& 6035 & L & $-$530.70 & $-$1766.70 & $-$44.44 &   6.53 &  0.172 &    34 & 20\\
 62& 6035 & R & $-$529.28 & $-$1767.08 & $-$44.05 &   2.18 &  0.135 &  $-$2 & 20\\
 63& 6030 & L & $-$529.16 & $-$1767.49 & $-$44.44 &   3.09 &  0.112 &    81 & 20\\
 64& 6035 & R & $-$528.58 & $-$1758.56 & $-$43.33 &   0.79 &  0.137 &    60 & \\
 65& 6030 & R & $-$528.29 & $-$1767.85 & $-$43.95 &   1.17 &  0.170 &    19 & 20\\
 66& 6030 & R & $-$525.53 & $-$1758.23 & $-$43.13 &   0.20 &  \nogradient   & \\
 67& 6035 & L & $-$515.96 & $-$1772.32 & $-$42.84 &   7.53 &  0.085 &    94 & 21\\
 68& 6035 & R & $-$515.79 & $-$1772.14 & $-$42.45 &   5.26 &  0.081 &   102 & 21\\
 69& 6030 & R & $-$512.17 & $-$1754.39 & $-$43.00 &   0.24 &  \nogradient   & (22)\\
 70& 6035 & R & $-$510.94 & $-$1754.40 & $-$43.08 &   0.98 &  \nogradient   & 22\\
 71& 6035 & L & $-$510.38 & $-$1755.02 & $-$43.42 &   1.61 &  0.166 &$-$176 & 22\\
 72& 6035 & R & $-$508.18 & $-$1778.81 & $-$43.98 &   0.37 &  \nogradient   & (23)\\
 73& 6030 & R & $-$506.30 & $-$1779.79 & $-$43.88 &   0.27 &  0.079 &    97 & 23\\
 74& 6030 & L & $-$505.99 & $-$1779.38 & $-$44.27 &   0.28 &  \nogradient   & 23\\
 75& 6030 & L & $-$359.01 & $-$1858.15 & $-$43.13 &   0.33 &  \nogradient   & 24\\
 76& 6035 & R & $-$357.05 & $-$1864.88 & $-$43.01 &   7.18 &  0.099 & $-$80 & 24\\
 77& 6035 & L & $-$356.89 & $-$1864.87 & $-$43.35 &   7.89 &  0.097 & $-$90 & 24\\
 78& 6030 & R & $-$356.77 & $-$1865.14 & $-$42.96 &   0.78 &  0.094 & $-$86 & 24\\
 79& 6035 & R & $-$318.30 & $-$1066.17 & $-$43.98 &   1.26 &  0.095 &   173 & \\
 80& 6035 & L & $-$309.04 & $-$1850.43 & $-$46.96 &   0.19 &  \nogradient   & 25\\
 81& 6035 & R & $-$308.72 & $-$1849.76 & $-$46.76 &   0.18 &  \nogradient   & 25\\
 82& 6035 & R & $-$299.93 & $-$1403.12 & $-$45.33 &   0.26 &  \nogradient   & 26\\
 83& 6035 & L & $-$299.30 & $-$1402.72 & $-$45.60 &   0.33 &  \nogradient   & 26\\
 84& 6030 & L & $-$293.62 & $-$1768.79 & $-$43.37 &   1.14 &  \nogradient   & 27\\
 85& 6030 & R & $-$292.96 & $-$1769.20 & $-$42.93 &   0.70 &  \nogradient   & 27\\
 86& 6035 & L & $-$292.94 & $-$1768.87 & $-$43.32 &   2.53 &  \nogradient   & 27\\
 87& 6035 & R & $-$292.78 & $-$1767.61 & $-$43.01 &   2.25 &  \nogradient   & 27\\
 88& 6035 & R & $-$289.42 & $-$1412.73 & $-$45.48 &   0.16 &  \nogradient   & \\
 89& 6035 & R & $-$271.21 &  $-$685.82 & $-$44.27 &   0.51 &  \nogradient   & 28\\
 90& 6035 & L & $-$270.94 &  $-$686.00 & $-$44.63 &   0.46 &  \nogradient   & 28\\
 91& 6035 & L & $-$244.91 &  $-$370.20 & $-$45.14 &   5.65 &  0.071 &$-$105 & 29\\
 92& 6035 & R & $-$244.23 &  $-$677.89 & $-$43.25 &   0.46 &  \nogradient   & \\
 93& 6035 & R & $-$242.53 &  $-$369.44 & $-$45.02 &  12.94 &  0.100 & $-$95 & 29\\
 94& 6035 & R & $-$241.80 & $-$1754.60 & $-$42.59 &   1.93 &  \nogradient   & (30)\\
 95& 6030 & R & $-$241.75 &  $-$368.97 & $-$45.00 &   0.28 &  0.279 & $-$95 & (29)\\
 96& 6030 & L & $-$241.04 & $-$1754.67 & $-$42.96 &   1.11 &  \nogradient   & 30\\
 97& 6030 & R & $-$240.54 & $-$1754.94 & $-$42.47 &   0.72 &  \nogradient   & 30\\
 98& 6035 & L & $-$239.57 &  $-$709.30 & $-$43.42 &   1.03 &  \nogradient   & \\
 99& 6035 & L & $-$211.45 &  $-$354.84 & $-$44.24 &   0.46 &  \nogradient   & 31\\
100& 6035 & R & $-$211.44 &  $-$355.35 & $-$43.88 &   0.35 &  \nogradient   & 31\\
101& 6035 & L & $-$208.70 & $-$1389.00 & $-$45.26 &   0.34 &  0.052 &    88 & \\
102& 6035 & L & $-$183.40 & $-$1184.30 & $-$43.93 &   0.60 &  \nogradient   & 32\\
103& 6035 & R & $-$183.38 &   $-$27.81 & $-$43.56 &   4.41 &  1.010 &$-$136 & 33\\
104& 6035 & L & $-$183.28 &   $-$27.77 & $-$43.98 &   5.38 &  1.087 & $-$84 & 33\\
105& 6035 & R & $-$183.24 & $-$1184.63 & $-$43.71 &   0.66 &  \nogradient   & 32\\
106& 6030 & L & $-$183.23 &   $-$28.12 & $-$44.05 &   1.02 &  0.455 &$-$169 & 33\\
107& 6030 & R & $-$183.21 &   $-$28.08 & $-$43.42 &   0.72 &  0.388 &   173 & 33\\
108& 6035 & R & $-$182.57 & $-$1386.07 & $-$44.46 &   1.35 &  1.587 &$-$116 & 34\\
109& 6035 & L & $-$182.30 & $-$1386.02 & $-$44.73 &   1.60 &  0.800 &   163 & 34\\
110& 6035 & L & $-$182.25 & $-$1174.33 & $-$43.95 &   0.77 &  0.055 &    18 & 35\\
111& 6035 & R & $-$181.60 & $-$1176.16 & $-$43.66 &   0.40 &  \nogradient   & 35\\
112& 6030 & R & $-$179.69 & $-$1386.84 & $-$44.36 &   0.14 &  \nogradient   & 34\\
113& 6030 & R & $-$179.60 & $-$1122.78 & $-$44.07 &   0.47 &  0.098 &   146 & 36\\
114& 6030 & L & $-$179.15 & $-$1122.83 & $-$44.36 &   0.43 &  0.066 &   148 & 36\\
115& 6030 & L & $-$178.83 & $-$1386.99 & $-$44.80 &   0.12 &  \nogradient   & 34\\
116& 6035 & R & $-$178.77 & $-$1122.31 & $-$44.12 &  17.25 &  0.060 &   162 & 36\\
117& 6035 & L & $-$178.34 & $-$1122.94 & $-$44.29 &  22.51 &  0.063 &   160 & 36\\
118& 6030 & R & $-$173.59 & $-$1762.46 & $-$41.94 &   0.20 &  \nogradient   & 37\\
119& 6030 & L & $-$172.52 & $-$1761.89 & $-$42.28 &   0.31 &  0.078 &    68 & 37\\
120& 6035 & L & $-$170.85 & $-$1761.32 & $-$42.18 &   2.89 &  0.097 &    72 & 37\\
121& 6035 & R & $-$170.78 & $-$1761.40 & $-$41.89 &   1.79 &  0.107 &    74 & 37\\
122& 6035 & L & $-$167.17 & $-$1152.70 & $-$43.05 &   2.44 &  \nogradient   & 38\\
123& 6035 & R & $-$166.86 & $-$1152.38 & $-$42.76 &   2.06 &  0.154 &    40 & 38\\
124& 6035 & L & $-$165.08 &   $-$45.43 & $-$43.68 &   0.46 &  \nogradient   & \\
125& 6035 & R & $-$161.13 & $-$1167.17 & $-$42.91 &   6.03 &  0.184 & $-$80 & 39\\
126& 6035 & L & $-$160.60 & $-$1166.61 & $-$43.22 &   7.35 &  0.175 & $-$92 & 39\\
127& 6035 & L & $-$159.19 &    $-$2.74 & $-$46.16 &   0.19 &  \nogradient   & 40\\
128& 6035 & R & $-$159.08 &    $-$2.35 & $-$45.77 &   0.22 &  0.181 &    19 & 40\\
129& 6030 & R & $-$156.21 & $-$1163.50 & $-$43.30 &   3.00 &  0.885 & $-$95 & 41\\
130& 6035 & R & $-$156.14 & $-$1163.32 & $-$43.37 &   5.80 &  0.270 &   175 & 41\\
131& 6030 & L & $-$156.12 & $-$1163.47 & $-$43.78 &   2.19 &  0.270 &$-$162 & 41\\
132& 6035 & L & $-$156.03 & $-$1163.24 & $-$43.71 &   5.21 &  0.235 &   179 & 41\\
133& 6030 & L & $-$155.53 & $-$1176.88 & $-$43.25 &   0.56 &  \nogradient   & 42\\
134& 6030 & R & $-$155.18 & $-$1176.02 & $-$42.83 &   0.80 &  0.076 &$-$165 & 42\\
135& 6035 & R & $-$153.74 &  $-$275.88 & $-$43.95 &   0.23 &  \nogradient   & \\
136& 6035 & R & $-$150.48 &       9.85 & $-$44.92 &   7.66 &  0.926 &    44 & 43\\
137& 6035 & L & $-$150.41 &       9.88 & $-$45.24 &   5.57 &  0.680 &    72 & 43\\
138& 6035 & L & $-$139.72 & $-$1756.56 & $-$41.55 &   0.14 &  \nogradient   & 44\\
139& 6035 & R & $-$139.54 & $-$1757.73 & $-$41.26 &   0.13 &  \nogradient   & 44\\
140& 6030 & R & $-$128.58 & $-$1315.41 & $-$43.37 &   0.57 &  0.315 &    51 & 45\\
141& 6030 & L & $-$128.40 & $-$1315.31 & $-$43.83 &   0.62 &  \nogradient   & 45\\
142& 6035 & R & $-$128.06 & $-$1315.13 & $-$43.44 &   2.40 &  0.099 &    65 & 45\\
143& 6035 & L & $-$128.04 & $-$1315.14 & $-$43.78 &   3.02 &  0.116 &    70 & 45\\
144& 6035 & L & $-$117.50 & $-$1352.66 & $-$43.73 &   1.91 &  0.543 &    85 & 46\\
145& 6035 & R & $-$117.50 & $-$1352.83 & $-$43.39 &   1.49 &  0.237 &    43 & 46\\
146& 6030 & R & $-$111.02 &   $-$45.72 & $-$43.20 &   0.55 &  0.271 &$-$148 & 47\\
147& 6030 & L & $-$110.99 &   $-$45.89 & $-$43.85 &   0.52 &  \nogradient   & 47\\
148& 6035 & R & $-$110.64 &   $-$45.96 & $-$43.39 &   3.58 &  0.521 &   163 & 47\\
149& 6035 & L & $-$110.60 &   $-$45.84 & $-$43.83 &   4.74 &  1.408 &    84 & 47\\
150& 6035 & L & $-$110.14 &      43.63 & $-$47.20 &   0.71 &  0.264 &$-$120 & \\
151& 6035 & R & $-$110.14 &      44.01 & $-$47.25 &   1.16 &  0.209 &$-$139 & \\
152& 6035 & L & $-$105.61 &   $-$51.80 & $-$43.59 &   2.38 &  0.093 & $-$28 & \\
153& 6035 & R & $-$102.08 & $-$1753.03 & $-$41.31 &   0.19 &  0.155 &   137 & 48\\
154& 6035 & L & $-$101.78 & $-$1752.70 & $-$41.55 &   0.36 &  0.269 &   118 & 48\\
155& 6035 & L &  $-$95.02 &   $-$99.56 & $-$43.18 &  16.24 &  0.180 & $-$24 & 49\\
156& 6035 & R &  $-$95.00 &   $-$99.53 & $-$42.74 &  10.14 &  0.172 & $-$39 & 49\\
157& 6030 & L &  $-$91.12 &  $-$104.81 & $-$43.47 &   2.42 &  \nogradient   & \\
158& 6035 & R &  $-$90.58 & $-$1387.48 & $-$43.47 &   0.93 &  \nogradient   & 50\\
159& 6035 & L &  $-$90.55 & $-$1387.62 & $-$43.81 &   1.36 &  0.152 &$-$145 & 50\\
160& 6035 & R &  $-$81.66 &   $-$62.76 & $-$43.32 &   1.07 &  \nogradient   & 51\\
161& 6035 & L &  $-$81.50 &   $-$62.94 & $-$43.81 &   0.82 &  \nogradient   & 51\\
162& 6030 & L &  $-$77.40 &  $-$171.91 & $-$42.76 &   2.23 &  0.289 &   178 & 52\\
163& 6030 & R &  $-$77.24 &  $-$172.01 & $-$42.25 &   4.67 &  0.187 & $-$66 & 52\\
164& 6035 & R &  $-$73.80 &  $-$172.44 & $-$42.40 &  15.66 &  0.201 & $-$78 & 52\\
165& 6035 & L &  $-$73.53 &  $-$172.36 & $-$42.79 &   9.00 &  0.268 & $-$52 & 52\\
166& 6035 & R &  $-$69.43 &   $-$74.55 & $-$42.57 & 121.57 &  1.408 & $-$89 & 53\\
167& 6035 & L &  $-$69.38 &   $-$74.52 & $-$43.27 & 129.96 &  0.909 & $-$87 & 53\\
168\tnm{f}& 6030 & R &  $-$68.83 &   $-$74.79 & $-$42.52 &  84.81 &  0.372 & $-$91 & 53\\
169& 6030 & L &  $-$67.96 &   $-$75.24 & $-$43.56 &  98.45 &  0.143 & $-$69 & 53\\
170& 6030 & L &  $-$67.79 &   $-$73.59 & $-$42.59 &   0.23 &  \nogradient   & \\
171& 6035 & R &  $-$65.61 &  $-$175.58 & $-$42.62 &   7.11 &  0.107 &$-$102 & \\
172& 6035 & R &  $-$65.00 &   $-$81.05 & $-$44.10 &   1.06 &  0.201 &    34 & 54\\
173& 6030 & R &  $-$64.73 &  $-$175.47 & $-$42.71 &  11.98 &  0.288 &$-$109 & 55\\
174& 6035 & L &  $-$64.57 &   $-$80.37 & $-$44.56 &   2.21 &  0.198 &    53 & 54\\
175& 6030 & L &  $-$64.48 &  $-$175.41 & $-$43.37 &  27.03 &  0.200 &$-$102 & 55\\
176& 6035 & L &  $-$64.46 &  $-$175.00 & $-$43.39 &  98.27 &  0.268 & $-$80 & 55\\
177& 6035 & R &  $-$64.31 &  $-$174.85 & $-$42.96 &  30.75 &  0.263 & $-$55 & 55\\
178& 6030 & R &  $-$63.39 &   $-$65.49 & $-$44.75 &   0.24 &  \nogradient   & (56)\\
179& 6035 & R &  $-$63.37 &   $-$79.32 & $-$43.78 &   0.90 &  0.189 &   152 & \\
180& 6035 & R &  $-$62.97 &   $-$65.14 & $-$44.90 &   2.37 &  0.621 &$-$120 & 56\\
181& 6035 & L &  $-$62.92 &   $-$64.70 & $-$45.48 &   0.43 &  \nogradient   & 56\\
182& 6035 & R &  $-$62.82 &   $-$63.99 & $-$45.84 &   0.11 &  \nogradient   & \\
183& 6035 & L &  $-$60.95 & $-$1732.94 & $-$42.30 &   0.37 &  0.136 &   110 & 57\\
184& 6035 & R &  $-$60.73 & $-$1733.89 & $-$42.04 &   0.40 &  \nogradient   & 57\\
185& 6035 & L &  $-$60.38 &     450.87 & $-$47.83 &   0.18 &  \nogradient   & 59\\
186& 6035 & L &  $-$60.20 &   $-$34.76 & $-$47.08 &   0.27 &  \nogradient   & 58\\
187& 6035 & R &  $-$60.12 &   $-$34.98 & $-$46.47 &   0.27 &  0.248 &    68 & 58\\
188& 6035 & R &  $-$59.83 &     450.61 & $-$47.59 &   0.20 &  \nogradient   & 59\\
189& 6035 & L &  $-$58.28 &  $-$175.23 & $-$43.56 &  11.12 &  0.610 &   146 & 60\\
190& 6035 & R &  $-$57.87 &  $-$175.28 & $-$43.15 &   5.78 &  0.068 & $-$94 & 60\\
191& 6035 & L &  $-$56.33 &   $-$82.18 & $-$45.02 &   0.35 &  \nogradient   & 61\\
192& 6035 & R &  $-$56.12 &   $-$81.70 & $-$44.32 &   0.29 &  \nogradient   & 61\\
193& 6035 & L &  $-$56.08 &      69.80 & $-$47.10 &   0.29 &  \nogradient   & \\
194& 6030 & L &  $-$55.91 &   $-$82.45 & $-$45.17 &   0.33 &  0.282 &    48 & 61\\
195& 6030 & R &  $-$54.97 &  $-$175.54 & $-$42.93 &   1.93 &  0.469 &$-$135 & \\
196& 6035 & R &  $-$54.33 &      69.92 & $-$47.10 &   0.15 &  \nogradient   & \\
197& 6030 & R &  $-$52.56 &   $-$81.47 & $-$43.85 &   1.66 &  0.917 &    94 & 61\\
198& 6035 & R &  $-$47.79 & $-$1728.06 & $-$42.50 &  13.33 &  0.285 &    98 & 62\\
199& 6035 & L &  $-$47.17 & $-$1727.91 & $-$42.79 &  10.73 &  0.403 &   133 & 62\\
200& 6030 & R &  $-$45.84 &  $-$103.82 & $-$44.00 &   0.18 &  \nogradient   & 63\\
201& 6030 & L &  $-$44.99 &  $-$103.14 & $-$44.80 &   0.17 &  \nogradient   & 63\\
202& 6035 & R &  $-$44.84 &  $-$102.86 & $-$44.34 &   0.28 &  \nogradient   & \\
203& 6035 & R &  $-$43.26 &  $-$169.62 & $-$43.35 &   4.84 &  1.136 &$-$130 & 64\\
204& 6035 & L &  $-$42.89 &  $-$169.42 & $-$43.78 &   2.02 &  0.296 &$-$109 & 64\\
205& 6030 & R &  $-$42.77 &  $-$169.70 & $-$43.32 &   4.24 &  0.422 &$-$107 & 64\\
206& 6030 & L &  $-$42.48 &  $-$169.68 & $-$43.95 &   1.60 &  0.442 & $-$58 & 64\\
207& 6035 & R &  $-$40.52 &      63.36 & $-$47.81 &   1.20 &  0.109 & $-$31 & 65\\
208& 6035 & R &  $-$39.73 &  $-$103.48 & $-$43.71 &   3.84 &  0.140 &    99 & 66\\
209& 6035 & L &  $-$38.95 &      59.75 & $-$48.22 &   0.52 &  0.059 & $-$46 & 65\\
210& 6030 & R &  $-$37.83 &  $-$104.15 & $-$43.51 &   0.71 &  0.124 &   118 & 66\\
211& 6030 & L &  $-$37.53 &  $-$104.23 & $-$44.36 &   0.77 &  0.134 &   108 & 66\\
212& 6035 & L &  $-$37.04 &  $-$104.02 & $-$44.07 &   5.43 &  0.133 &   102 & 66\\
213& 6030 & L &  $-$36.55 &   $-$10.61 & $-$46.50 &   0.50 &  0.179 &    66 & 67\\
214& 6030 & R &  $-$36.20 &   $-$10.84 & $-$45.58 &   0.49 &  0.244 &    75 & 67\\
215& 6035 & R &  $-$35.76 &    $-$9.87 & $-$45.63 &   0.39 &  0.245 &    99 & 67\\
216& 6035 & L &  $-$35.37 &    $-$9.83 & $-$46.26 &   0.44 &  0.164 &    80 & 67\\
217& 6030 & L &  $-$34.27 &   $-$66.98 & $-$42.86 &   4.24 &  0.746 &    29 & \\
218& 6035 & L &  $-$31.84 & $-$1767.69 & $-$45.12 &   0.19 &  \nogradient   & \\
219& 6035 & L &  $-$28.07 &       4.65 & $-$46.86 &   0.24 &  0.275 &$-$117 & \\
220& 6030 & L &  $-$26.66 &    $-$0.54 & $-$45.41 &   2.30 &  1.282 &   144 & 68\\
221& 6030 & R &  $-$26.66 &    $-$0.53 & $-$44.56 &   2.88 &  1.613 &$-$176 & 68\\
222& 6035 & R &  $-$26.46 &    $-$0.10 & $-$44.68 &   7.25 &  0.943 &   142 & 68\\
223& 6035 & L &  $-$26.32 &    $-$0.11 & $-$45.26 &   5.98 &  1.818 &   161 & 68\\
224& 6035 & R &  $-$23.58 &      13.56 & $-$45.46 &   0.67 &  0.699 &$-$133 & 69\\
225& 6035 & L &  $-$23.56 &      13.84 & $-$45.99 &   1.00 &  0.261 &   151 & 69\\
226& 6035 & L &  $-$22.36 &    $-$3.79 & $-$44.87 &   0.24 &  \nogradient   & \\
227& 6035 & R &  $-$17.88 &      26.62 & $-$44.90 &   3.30 &  0.255 &    56 & 70\\
228& 6035 & L &  $-$17.67 &      26.73 & $-$45.26 &   4.40 &  0.186 &    43 & 70\\
229& 6030 & R &  $-$14.01 &  $-$109.91 & $-$42.08 &   0.61 &  \nogradient   & \\
230& 6035 & R &  $-$12.50 &      31.45 & $-$44.58 &   0.69 &  0.062 &    40 & \\
231& 6030 & L &  $-$11.25 &      30.10 & $-$42.83 &   6.41 &  0.641 &   111 & \\
232& 6030 & L &  $-$10.56 &       5.86 & $-$45.65 &   0.30 &  0.128 &   111 & (71)\\
233& 6035 & R &  $-$10.10 &       5.92 & $-$45.21 &   7.25 &  0.172 &   127 & 71\\
234& 6035 & L &   $-$9.87 &       5.87 & $-$45.60 &  51.48 &  0.163 &   124 & 71\\
235& 6030 & R &   $-$8.94 &  $-$116.47 & $-$41.84 &   1.79 &  0.350 &   169 & 73\\
236& 6035 & L &   $-$8.94 &      37.00 & $-$44.61 &   0.85 &  0.185 &    65 & 72\\
237& 6035 & R &   $-$8.94 &      36.37 & $-$44.29 &   0.37 &  \nogradient   & 72\\
238& 6030 & L &   $-$8.80 &  $-$116.69 & $-$42.69 &   0.79 &  \nogradient   & 73\\
239& 6035 & L &   $-$8.54 &  $-$115.88 & $-$42.45 &   7.31 &  0.260 &$-$167 & 73\\
240& 6035 & R &   $-$8.41 &  $-$116.10 & $-$41.82 &   5.87 &  0.259 &$-$122 & 73\\
241& 6035 & R &   $-$8.19 &   $-$96.64 & $-$41.94 &  27.82 &  0.122 &     6 & 74\\
242& 6035 & L &   $-$8.08 &   $-$96.61 & $-$42.57 &  18.76 &  0.193 &    12 & 74\\
243& 6030 & R &   $-$7.86 &   $-$94.31 & $-$41.86 &   4.81 &  0.493 &     0 & 74\\
244& 6030 & L &   $-$7.76 &   $-$95.66 & $-$42.71 &   1.61 &  \nogradient   & 74\\
245& 6035 & L &   $-$7.62 &   $-$80.09 & $-$42.11 &   0.17 &  \nogradient   & \\
246& 6035 & R &   $-$7.23 &   $-$80.59 & $-$42.11 &  45.55 &  0.183 &$-$160 & 75\\
247& 6035 & L &   $-$7.04 &   $-$80.45 & $-$42.79 &  51.03 &  0.360 &$-$161 & 75\\
248& 6030 & R &   $-$6.47 &   $-$78.72 & $-$42.03 &   8.90 &  0.336 &$-$163 & 75\\
249& 6030 & L &   $-$5.75 &   $-$78.45 & $-$42.88 &   7.00 &  0.397 &    37 & 75\\
250& 6035 & L &   $-$5.13 &     413.95 & $-$47.37 &   0.45 &  0.106 &$-$156 & 76\\
251& 6030 & R &   $-$0.66 &    $-$0.23 & $-$42.83 &   9.84 &  0.427 &    77 & \\
252& 6035 & L &   $-$0.19 &    $-$0.01 & $-$43.78 &   2.59 &  1.887 &$-$165 & \\
253& 6035 & R &   $-$0.14 &    $-$0.07 & $-$42.93 & 251.49 &  0.855 &    86 & \\
254& 6030 & L &   $-$0.08 &    $-$0.38 & $-$42.86 & 138.79 &  1.031 &   101 & \\
255& 6035 & R &   $-$0.05 &     415.44 & $-$47.30 &   0.30 &  \nogradient   & 76\\
256\tnm{f}& 6035 & L &      0.03 &    $-$0.01 & $-$42.93 & 576.47 &  1.695 &    78 & \\
257& 6030 & R &      4.57 &     416.05 & $-$47.37 &   0.55 &  0.247 &$-$156 & (77)\\
258& 6035 & L &      6.24 &     415.58 & $-$47.64 &   0.34 &  0.134 &   172 & 77\\
259& 6030 & R &      6.52 & $-$1704.86 & $-$43.51 &   1.01 &  0.310 & $-$29 & 78\\
260& 6035 & R &      6.65 &     416.28 & $-$47.44 &   0.93 &  \nogradient   & 77\\
261& 6030 & L &      6.76 & $-$1704.91 & $-$43.85 &   0.94 &  \nogradient   & 78\\
262& 6035 & R &      6.79 & $-$1704.42 & $-$43.64 &   0.87 &  0.654 &     2 & 78\\
263& 6035 & L &      6.98 & $-$1704.34 & $-$43.85 &   1.11 &  0.412 & $-$26 & 78\\
264& 6035 & R &      7.66 &     414.24 & $-$47.42 &   0.52 &  \nogradient   & 79\\
265& 6035 & L &      9.99 &     415.10 & $-$47.76 &   0.24 &  \nogradient   & 79\\
266& 6035 & L &     11.03 &      48.34 & $-$43.32 &   1.90 &  \nogradient   & \\
267& 6030 & R &     11.31 &   $-$15.81 & $-$42.93 &   0.78 &  0.143 & $-$35 & (g)\\
268& 6035 & L &     11.32 &   $-$16.83 & $-$43.49 &   3.14 &  0.046 &   179 & (g)\\
269& 6035 & L &     22.81 & $-$1708.61 & $-$41.96 &   0.12 &  \nogradient   & \\
270& 6035 & L &     84.28 & $-$1755.55 & $-$44.61 &   0.54 &  \nogradient   & 80\\
271& 6035 & R &     84.87 & $-$1755.84 & $-$44.34 &   0.45 &  \nogradient   & 80\\
272& 6030 & L &    176.67 & $-$1781.90 & $-$42.98 &   0.81 &  \nogradient   & 81\\
273& 6030 & R &    176.83 & $-$1781.99 & $-$42.67 &   0.44 &  \nogradient   & 81\\
274& 6035 & L &    195.67 & $-$1783.93 & $-$43.59 &   0.73 &  \nogradient   & 82\\
275& 6035 & R &    195.72 & $-$1783.98 & $-$43.32 &   0.56 &  \nogradient   & 82\\
276& 6035 & L &    212.04 & $-$1743.12 & $-$43.88 &   0.29 &  \nogradient   & 83\\
277& 6035 & R &    212.33 & $-$1744.07 & $-$43.69 &   0.25 &  \nogradient   & 83\\
278& 6035 & R &    387.46 & $-$1767.82 & $-$48.08 &   0.34 &  0.124 & $-$11 & 84\\
279& 6035 & L &    388.00 & $-$1768.12 & $-$48.00 &   0.49 &  0.155 & $-$44 & 84\\
280& 6035 & R &    634.06 & $-$1592.53 & $-$47.78 &   0.57 &  0.210 & $-$74 & 85\\
281& 6035 & L &    634.11 & $-$1592.40 & $-$47.74 &   1.58 &  0.230 & $-$69 & 85\\
282& 6035 & R &    701.99 & $-$1250.76 & $-$48.24 &   0.23 &  5.000 &$-$150 & \\
283& 6035 & L &    702.02 & $-$1250.65 & $-$48.24 &   0.40 &  0.223 &$-$138 & \\
284& 6035 & L &    719.64 & $-$1240.55 & $-$48.05 &   0.50 &  0.368 &    94 & \\
285& 6035 & L &   1060.93 &     398.63 & $-$49.21 &   0.23 &  \nogradient   & 86\\
286& 6035 & R &   1061.36 &     398.23 & $-$49.14 &   0.19 &  \nogradient   & 86\\
287& 6035 & L &   1218.62 &     321.27 & $-$48.63 &   0.17 &  \nogradient   & 87\\
288& 6035 & R &   1219.96 &     319.90 & $-$48.68 &   0.16 &  \nogradient   & 87\\
289& 6035 & R &   1225.53 &     309.60 & $-$48.83 &   2.42 &  1.176 & $-$29 & 88\\
290& 6035 & L &   1225.58 &     309.69 & $-$48.80 &   2.69 &  1.000 & $-$62 & 88\\
291& 6035 & R &   1426.03 &     103.76 & $-$49.60 &   0.21 &  \nogradient   & 89\\
292& 6035 & L &   1426.08 &     104.63 & $-$49.53 &   0.20 &  \nogradient   & 89\\
\enddata
%
%
\notetoeditor{This may be best as an online-only table with a stub 10-row
              table in the print edition.}
\tnt{a}{Centroid of Gaussian fit in channel of peak emission.  RA and
  Declination offsets are with respect to
  $02^\mathrm{h}27^\mathrm{m}03\fs8343, +61\degr52\arcmin25\farcs300
  \pm 0\farcs01$.  Errors range from $< 0.01$~mas for the strongest
  maser spots to $\lesssim 1$~mas for the weakest detections.}
\tnt{b}{LSR velocity of channel of peak emission.}
\tnt{c}{Peak brightness of Gaussian fit in channel of peak emission.}
\tnt{d}{Position angle east of north in direction of increasing
        line$-$of$-$sight velocity across the maser spot.}
\tnt{e}{See Table \ref{tab-zeeman}.  Parenthesized numbers indicate unpaired
        masers in one transition consistent with the central velocity and
        magnetic field strength derived from a Zeeman pair in the other
        transition.}
\tnt{f}{Reference feature.}
\tnt{g}{Consistent with $B = 8.3$~mG, $v_\mathrm{LSR} = -43.26$~km~s$^{-1}$
        if interpreted as a multi-transition Zeeman association.}
\end{deluxetable}

\begin{deluxetable}{lrrcrrr}
\tabletypesize{\small}
\tablecaption{Zeeman Pairs and Associations\label{tab-zeeman}}
\tablehead{
  \colhead{Group} &
  \colhead{R.A.~Offset\tnm{a}} &
  \colhead{Decl.~Offset\tnm{a}} &
  \colhead{Freq.} &
  \colhead{Velocity\tnm{b}} &
  \colhead{$B$\tnm{c}} &
  \colhead{Separation} \\
  \colhead{Number} &
  \colhead{(mas)} &
  \colhead{(mas)} &
  \colhead{(MHz)} &
  \colhead{(km\,s$^{-1}$)} &
  \colhead{(mG)} & 
  \colhead{(mas)}
}
\startdata
 1& $-$972.84  &     49.49  & 6035 &$-$46.27  &   4.7  & 2.9 \\
 2& $-$961.74  &     40.35  & 6035 &$-$46.18  &   5.2  & 0.3 \\
 3& $-$942.32  &    158.84  & 6035 &$-$46.66  &   5.6  & 1.4 \\
 4& $-$934.38  &    169.25  & 6035 &$-$47.00  &   5.6  & 0.2 \\
 5& $-$930.58  &    143.36  & 6035 &$-$47.18  &   6.0  & 0.2 \\
  & $-$930.44  &    142.61  & 6030 &$-$47.11  &   6.2  & 0.1 \\
 6& $-$926.10  &    128.09  & 6035 &$-$46.72  &   5.2  & 1.2 \\
 7& $-$918.27  &     59.06  & 6035 &$-$45.44  &   6.5  & 0.1 \\
  & $-$916.70  &     60.34  & 6030 &$-$45.65  &   6.1  & 0.1 \\
 8& $-$896.50  &     82.29  & 6030 &$-$45.47  &   5.8  & 0.1 \\
  & $-$896.38  &     82.49  & 6035 &$-$45.38  &   6.0  & 0.2 \\
 9& $-$867.20  & $-$607.07  & 6035 &$-$44.96  &   3.0  & 0.0 \\
10& $-$865.84  & $-$641.56  & 6030 &$-$45.01  &   2.2  & 4.4 \\
11& $-$705.08  &$-$1871.16  & 6030 &$-$43.08  &   6.1  & 0.3 \\
  & $-$705.04  &$-$1872.15  & 6035 &$-$43.11  &   5.6  & 1.2 \\
12& $-$698.04  &$-$1874.31  & 6030 &$-$43.03  &   6.2  & 3.3 \\
  & $-$692.40  &$-$1879.39  & 6035 &$-$43.12  &   6.5  & 0.4 \\
13& $-$679.09  &$-$1892.51  & 6030 &$-$42.97  &   6.4  & 1.5 \\
14& $-$668.94  &$-$1898.89  & 6035 &$-$43.22  &   6.9  & 0.7 \\
15& $-$612.22  &$-$1907.35  & 6035 &$-$43.47  &   6.0  & 0.7 \\
16& $-$606.49  &$-$1907.77  & 6030 &$-$43.50  &   6.4  & 0.3 \\
  & $-$605.80  &$-$1907.28  & 6035 &$-$43.52  &   6.9  & 0.2 \\
17& $-$568.86  &$-$1742.78  & 6035 &$-$44.81  &   4.7  & 1.1 \\
18& $-$561.80  &$-$1729.99  & 6035 &$-$44.80  &   4.3  & 0.3 \\
19& $-$534.33  &$-$1765.18  & 6035 &$-$44.42  &   3.9  & 2.3 \\
20& $-$529.99  &$-$1766.89  & 6035 &$-$44.24  &   6.9  & 1.5 \\
  & $-$528.72  &$-$1767.67  & 6030 &$-$44.19  &   6.2  & 0.9 \\
21& $-$515.88  &$-$1772.23  & 6035 &$-$42.64  &   6.9  & 0.3 \\
22& $-$510.66  &$-$1754.71  & 6035 &$-$43.25  &   6.0  & 0.8 \\
23& $-$506.15  &$-$1779.58  & 6030 &$-$44.07  &   4.9  & 0.5 \\
24\tnm{d}& $-$357.89  &$-$1861.65  & 6030 &$-$43.04  &   2.2  & 7.3 \\
  & $-$356.97  &$-$1864.87  & 6035 &$-$43.18  &   6.0  & 0.2 \\
25& $-$308.88  &$-$1850.10  & 6035 &$-$46.86  &   3.4  & 0.7 \\
26& $-$299.61  &$-$1402.92  & 6035 &$-$45.47  &   4.7  & 0.7 \\
27& $-$293.29  &$-$1769.00  & 6030 &$-$43.15  &   5.5  & 0.8 \\
  & $-$292.86  &$-$1768.24  & 6035 &$-$43.16  &   5.6  & 1.3 \\
28& $-$271.07  & $-$685.91  & 6035 &$-$44.45  &   6.5  & 0.3 \\
29& $-$243.72  & $-$369.82  & 6035 &$-$45.08  &   2.1  & 2.5 \\
30& $-$240.79  &$-$1754.80  & 6030 &$-$42.71  &   6.1  & 0.6 \\
31& $-$211.45  & $-$355.09  & 6035 &$-$44.06  &   6.5  & 0.5 \\
32& $-$183.32  &$-$1184.47  & 6035 &$-$43.82  &   3.9  & 0.4 \\
33& $-$183.33  &  $-$27.79  & 6035 &$-$43.77  &   7.3  & 0.1 \\
  & $-$183.22  &  $-$28.10  & 6030 &$-$43.73  &   8.0  & 0.0 \\
34& $-$182.44  &$-$1386.04  & 6035 &$-$44.59  &   4.7  & 0.3 \\
  & $-$179.26  &$-$1386.92  & 6030 &$-$44.58  &   5.5  & 0.9 \\
35& $-$181.92  &$-$1175.25  & 6035 &$-$43.81  &   5.2  & 1.9 \\
36& $-$179.38  &$-$1122.80  & 6030 &$-$44.22  &   3.7  & 0.5 \\
  & $-$178.55  &$-$1122.63  & 6035 &$-$44.21  &   3.0  & 0.8 \\
37& $-$173.06  &$-$1762.18  & 6030 &$-$42.11  &   4.3  & 1.2 \\
  & $-$170.82  &$-$1761.36  & 6035 &$-$42.04  &   5.2  & 0.1 \\
38& $-$167.02  &$-$1152.54  & 6035 &$-$42.91  &   5.2  & 0.4 \\
39& $-$160.86  &$-$1166.89  & 6035 &$-$43.07  &   5.6  & 0.8 \\
40& $-$159.13  &   $-$2.55  & 6035 &$-$45.96  &   6.9  & 0.4 \\
41& $-$156.17  &$-$1163.49  & 6030 &$-$43.54  &   6.2  & 0.1 \\
  & $-$156.08  &$-$1163.28  & 6035 &$-$43.54  &   6.0  & 0.1 \\
42& $-$155.36  &$-$1176.45  & 6030 &$-$43.04  &   5.2  & 0.9 \\
43& $-$150.44  &      9.87  & 6035 &$-$45.08  &   5.6  & 0.1 \\
44& $-$139.63  &$-$1757.14  & 6035 &$-$41.40  &   5.2  & 1.2 \\
45& $-$128.49  &$-$1315.36  & 6030 &$-$43.60  &   5.8  & 0.2 \\
  & $-$128.05  &$-$1315.13  & 6035 &$-$43.61  &   6.0  & 0.0 \\
46& $-$117.50  &$-$1352.74  & 6035 &$-$43.56  &   6.0  & 0.2 \\
47& $-$111.00  &  $-$45.81  & 6030 &$-$43.53  &   8.3  & 0.2 \\
  & $-$110.62  &  $-$45.90  & 6035 &$-$43.61  &   7.7  & 0.1 \\
48& $-$101.93  &$-$1752.86  & 6035 &$-$41.43  &   4.3  & 0.4 \\
49&  $-$95.01  &  $-$99.55  & 6035 &$-$42.96  &   7.7  & 0.0 \\
50&  $-$90.56  &$-$1387.55  & 6035 &$-$43.64  &   6.0  & 0.1 \\
51&  $-$81.58  &  $-$62.85  & 6035 &$-$43.56  &   8.6  & 0.2 \\
52&  $-$77.32  & $-$171.96  & 6030 &$-$42.51  &   6.5  & 0.2 \\
  &  $-$73.66  & $-$172.40  & 6035 &$-$42.59  &   6.9  & 0.3 \\
53&  $-$69.40  &  $-$74.53  & 6035 &$-$42.92  &  12.5  & 0.1 \\
  &  $-$68.40  &  $-$75.01  & 6030 &$-$43.04  &  13.2  & 1.0 \\
54&  $-$64.78  &  $-$80.71  & 6035 &$-$44.33  &   8.2  & 0.8 \\
55&  $-$64.61  & $-$175.44  & 6030 &$-$43.04  &   8.3  & 0.3 \\
  &  $-$64.38  & $-$174.92  & 6035 &$-$43.18  &   7.7  & 0.2 \\
56&  $-$62.94  &  $-$64.92  & 6035 &$-$45.19  &  10.3  & 0.4 \\
57&  $-$60.84  &$-$1733.41  & 6035 &$-$42.17  &   4.7  & 1.0 \\
58&  $-$60.16  &  $-$34.87  & 6035 &$-$46.78  &  10.7  & 0.2 \\
59&  $-$60.10  &    450.74  & 6035 &$-$47.71  &   4.3  & 0.6 \\
60&  $-$58.08  & $-$175.25  & 6035 &$-$43.36  &   7.3  & 0.4 \\
61\tnm{d}&  $-$56.23  &  $-$81.94  & 6035 &$-$44.67  &  12.5  & 0.5 \\
  &  $-$54.23  &  $-$81.96  & 6030 &$-$44.51  &  16.6  & 3.5 \\
62&  $-$47.48  &$-$1727.98  & 6035 &$-$42.64  &   5.2  & 0.6 \\
63&  $-$45.42  & $-$103.48  & 6030 &$-$44.40  &  10.1  & 1.1 \\
64&  $-$43.08  & $-$169.52  & 6035 &$-$43.56  &   7.7  & 0.4 \\
  &  $-$42.63  & $-$169.69  & 6030 &$-$43.64  &   8.0  & 0.3 \\
65&  $-$39.74  &     61.55  & 6035 &$-$48.01  &   7.3  & 3.9 \\
66&  $-$38.38  & $-$103.75  & 6035 &$-$43.89  &   6.5  & 2.7 \\
  &  $-$37.68  & $-$104.19  & 6030 &$-$43.94  &  10.7  & 0.3 \\
67&  $-$36.37  &  $-$10.72  & 6030 &$-$46.04  &  11.7  & 0.4 \\
  &  $-$35.57  &   $-$9.85  & 6035 &$-$45.94  &  11.2  & 0.4 \\
68&  $-$26.65  &   $-$0.54  & 6030 &$-$44.98  &  10.8  & 0.0 \\
  &  $-$26.39  &   $-$0.11  & 6035 &$-$44.97  &  10.3  & 0.1 \\
69&  $-$23.57  &     13.70  & 6035 &$-$45.72  &   9.5  & 0.3 \\
70&  $-$17.78  &     26.68  & 6035 &$-$45.08  &   6.5  & 0.2 \\
71&   $-$9.98  &      5.89  & 6035 &$-$45.41  &   6.9  & 0.2 \\
72&   $-$8.94  &     36.69  & 6035 &$-$44.45  &   5.6  & 0.6 \\
73&   $-$8.87  & $-$116.58  & 6030 &$-$42.26  &  10.7  & 0.3 \\
  &   $-$8.48  & $-$115.99  & 6035 &$-$42.13  &  11.2  & 0.3 \\
74&   $-$8.13  &  $-$96.63  & 6035 &$-$42.25  &  11.2  & 0.1 \\
  &   $-$7.81  &  $-$94.98  & 6030 &$-$42.29  &  10.8  & 1.3 \\
75&   $-$7.14  &  $-$80.52  & 6035 &$-$42.45  &  12.0  & 0.2 \\
  &   $-$6.11  &  $-$78.59  & 6030 &$-$42.46  &  10.7  & 0.8 \\
76&   $-$2.59  &    414.69  & 6035 &$-$47.34  &   1.3  & 5.3 \\
77&      6.44  &    415.93  & 6035 &$-$47.54  &   3.4  & 0.8 \\
78&      6.88  &$-$1704.38  & 6035 &$-$43.74  &   3.9  & 0.2 \\
  &      6.64  &$-$1704.89  & 6030 &$-$43.68  &   4.3  & 0.2 \\
79&      8.82  &    414.67  & 6035 &$-$47.59  &   6.0  & 2.5 \\
80&     84.58  &$-$1755.70  & 6035 &$-$44.47  &   4.7  & 0.7 \\
81&    176.75  &$-$1781.94  & 6030 &$-$42.82  &   4.0  & 0.2 \\
82&    195.69  &$-$1783.96  & 6035 &$-$43.46  &   4.7  & 0.1 \\
83&    212.19  &$-$1743.59  & 6035 &$-$43.78  &   3.4  & 1.0 \\
84&    387.73  &$-$1767.97  & 6035 &$-$48.04  &$-$1.3  & 0.6 \\
85&    634.08  &$-$1592.47  & 6035 &$-$47.76  &$-$0.9  & 0.1 \\
86&   1061.15  &    398.43  & 6035 &$-$49.18  &   1.3  & 0.6 \\
87&   1219.29  &    320.59  & 6035 &$-$48.66  &$-$0.9  & 1.9 \\
88&   1225.55  &    309.64  & 6035 &$-$48.82  &$-$0.4  & 0.1 \\
89&   1426.06  &    104.19  & 6035 &$-$49.57  &$-$1.3  & 0.9 \\
\enddata
\tnt{a}{Average of the R.A.\ and Decl.\ offsets of the LCP and RCP
  components of Zeeman pair.}
\tnt{b}{Center LSR velocity of Zeeman pair.}
\tnt{c}{Positive values indicate magnetic fields oriented in the
        hemisphere pointing away from the observer.}
\tnt{d}{See \S \ref{zeeman} for discussion of this Zeeman
        association.}
\end{deluxetable}


\begin{thebibliography}{}

\bibitem[Alakoz et al.(2005)]{alakoz05} Alakoz, A.~V., Slysh, V.~I.,
  Popov, M.~V., \& Val'tts, I.~E.\ 2005, Astronomy Letters, 31, 375

\bibitem[A\u{\i}rapetyan et al.(1989)]{airapetyan89} A\u{\i}rapetyan,
  \'{E}.~A., Matveenko, L.~I., Kostenko, V.~I., Velikhov, V.~E.,
  Kopelyanski\u{\i}, G.~D., Molodyanu, A.~P., \& Timofeev, V.~V.\ 1989,
  \sovast~Lett., 15, 175

\bibitem[Baudry et al.(1997)]{baudry97} Baudry, A., Desmurs, J.~F.,
  Wilson, T.~L., \& Cohen, R.~J.\ 1997, \aap, 325, 255

\bibitem[Baudry \& Diamond(1991)]{baudry91} Baudry, A., \& Diamond,
  P.~J.\ 1991, \aap, 247, 551

\bibitem[Baudry \& Diamond(1998)]{baudry98} Baudry, A., \& Diamond,
  P.~J.\ 1998, \aap, 331, 697

\bibitem[Baudry et al.(1988)]{baudry88} Baudry, A., Diamond, P.~J.,
  Booth, R.~S., Graham, D., \& Walmsley, C.~M.\ 1988, \aap, 201, 105

\bibitem[Baudry et al.(1993)]{baudry93} Baudry, A., Menten, K.~M.,
  Walmsley, C.~M., \& Wilson, T.~L.\ 1993, \aap, 271, 552

\bibitem[Bloemhof et al.(1992)]{bloemhof92} Bloemhof, E.~E., Reid,
  M.~J., \& Moran, J.~M.\ 1992, \apj, 397, 500

\bibitem[B\"{o}ger et al.(2003)]{boger03} B\"{o}ger, R., Kegel,
  W.~H., \& Hegmann, M.\ 2003, \aap, 406, 23

\bibitem[Bourke et al.(2001)]{bourke01} Bourke, T.~L., Myers, P.~C.,
  Robinson, G., \& Hyland, A.~R.\ 2001, \apj, 554, 916

\bibitem[Caswell(2001)]{caswell01} Caswell, J.~L.\ 2001, \mnras, 326,
  805

\bibitem[Caswell(2003)]{caswell03} Caswell, J.~L.\ 2003, \mnras, 341,
  551

\bibitem[Caswell(2004)]{caswell04} Caswell, J.~L.\ 2004, \mnras, 352,
  101 

\bibitem[Cragg et al.(2002)]{cragg02} Cragg, D.~M., Sobolev, A.~M., \&
  Godfrey, P.~D.\ 2002, \mnras, 331, 521

\bibitem[De Buizer(2003)]{debuizer03} De Buizer, J.~M.\ 2003, \mnras,
  341, 277

\bibitem[Desmurs et al.(1998)]{desmurs98} Desmurs, J.~F., Baudry, A.,
  Wilson, T.~L., Cohen, R.~J., \& Tofani, G.\ 1998, \aap, 334, 1085
  (D98)

\bibitem[Dickel \& Goss(1987)]{dickel87} Dickel, H.~R., \& Goss,
  W.~M.\ 1987, \aap, 185, 271

\bibitem[Dodson et al.(2004)]{dodson04} Dodson, R., Ojha, R., \&
  Ellingsen, S.~P.\ 2004, \mnras, 351, 779

\bibitem[Elitzur(1994)]{elitzur94} Elitzur, M.\ 1994, \apj, 422, 751

\bibitem[Elitzur \& de Jong(1978)]{elitzur78} Elitzur, M., \& de Jong,
  T.\ 1976, \aap, 67, 323

\bibitem[Etoka et al.(2005)]{etoka05} Etoka, S., Cohen, R.~J., \&
  Gray, M.~D.\ 2005, \mnras, 360, 1162

\bibitem[Field et al.(1994)]{field94} Field, D., Gray, M.~D., \& de
  St.\ Paer, P.\ 1994, \aap, 282, 213

\bibitem[Fish(2007)]{fish07} Fish, V.~L.\ 2007, IAU Symposium 242, in
  press, arXiv:0704.0242

\bibitem[Fish et al.(2006a)]{fbs06} Fish, V.~L., Brisken, W.~F., \&
  Sjouwerman, L.~O.\ 2006a, \apj, 647, 418 (FBS)

\bibitem[Fish \& Reid(2006)]{fish06} Fish, V.~L., \& Reid, M.~J.\
  2006, \apjs, 164, 99

\bibitem[Fish et al.(2005)]{fish05} Fish, V.~L., Reid, M.~J., \&
  Menten, K.~M.\ 2005, \apj, 623, 269

\bibitem[Fish et al.(2006b)]{fisheff06} Fish, V.~L., Reid, M.~J.,
  Menten, K.~M., \& Pillai, T.\ 2006b, \aap, 458, 485

\bibitem[Fouquet \& Reid(1982)]{fouquet82} Fouquet, J.~E., \& Reid,
  M.~J.\ 1982, \aj, 87, 691

\bibitem[Fomalont et al.(2003)]{fomalont03} Fomalont, E.~B., Petrov,
  L., MacMillan, D.~S., Gordon, D., \& Ma, C.\ 2003, \aj, 126, 2562

\bibitem[Garay et al.(1989)]{garay89} Garay, G., Moran, J.~M., \&
  Haschick, A.~D.\ 1989, \apj, 338, 244

\bibitem[Garc\'{\i}a-Barreto et al.(1988)]{garciabarreto88}
  Garc\'{\i}a-Barreto, J.~A., Burke, B.~F., Reid, M.~J., Moran, J.~M.,
  Haschick, A.~D., \& Schilizzi, R.~T.\ 1988, \apj, 326, 954

\bibitem[Gray(2001)]{gray01} Gray, M.~D.\ 2001, \mnras, 324, 57

\bibitem[Gray et al.(1991)]{gray91} Gray, M.~D., Doel, R.~C., \&
  Field, D.\ 1991, \mnras, 252, 30

\bibitem[Guilloteau et al.(1985)]{guilloteau85} Guilloteau, S.,
  Baudry, A., \& Walmsley, C.~M.\ 1985, \aap, 153, 179

\bibitem[Guilloteau et al.(1983)]{guilloteau83} Guilloteau, S., Stier,
  M.~T., \& Downes, D.\ 1983, \aap, 126, 10

\bibitem[G\"{u}sten et al.(1994)]{gusten94} G\"{u}sten, R., Fiebig,
  D., \& Uchida, K.~I.\ 1994, \aap, 286, L51

\bibitem[Hachisuka et al.(2006)]{hachisuka06} Hachisuka, K., et al.\
  2006, \apj, 645, 337

\bibitem[Hansen et al.(1993)]{hansen93} Hansen, J., Booth, R.~S.,
  Dennison, B., \& Diamond, P.~J.\ 1993, LNP Vol.~412: Astrophysical
  Masers, 412, 255

\bibitem[Harvey-Smith \& Cohen(2005)]{harveysmith05} Harvey-Smith, L.,
  \& Cohen, R.~J.\ 2005, \mnras, 356, 637

\bibitem[Harvey-Smith \& Cohen(2006)]{harveysmith06} Harvey-Smith, L.,
  \& Cohen, R.~J.\ 2006, \mnras, 371, 1550

\bibitem[Hoffman et al.(2003)]{hoffman03} Hoffman, I.~M., Goss, W.~M.,
  Palmer, P., \& Richards, A.~M.~S.\ 2003, \apj, 598, 1061

\bibitem[Kawamura \& Masson(1998)]{kawamura98} Kawamura, J.~H., \&
  Masson, C.~R.\ 1998, \apj, 509, 270

\bibitem[Keto et al.(1995)]{keto95} Keto, E.~R., Welch, W.~J., Reid,
  M.~J., \& Ho, P.~T.~P.\ 1995, \apj, 444, 765

\bibitem[Lo et al.(1975)]{lo75} Lo, K.~Y., Walker, R.~C., Burke,
  B.~F., Moran, J.~M., Johnston, K.~J., \& Ewing, M.~S.\ 1975, \apj,
  202, 650

\bibitem[Mader et al.(1978)]{mader78} Mader, G.~L., Johnston, K.~J.,
  \& Moran, J.~M.\ 1978, \apj, 224, 115

\bibitem[Masheder et al.(1994)]{masheder94} Masheder, M.~R.~W., Field,
  D., Gray, M.~D., Migenes, V., Cohen, R.~J., \& Booth, R.~S.\ 1994,
  \aap, 281, 871

\bibitem[Menten et al.(1988)]{menten88} Menten, K.~M., Johnston,
  K.~J., Wadiak, E.~J., Walmsley, C.~M., \& Wilson, T.~L.\ 1988, \apj,
  331, L41

\bibitem[Moran et al.(1968)]{moran68} Moran, J.~M., Burke, B.~F.,
  Barrett, A.~H., Rogers, A.~E.~E., Ball, J.~A., Carter, J.~C., \&
  Cudaback, D.~D.\ 1968, \apj, 152, L97

\bibitem[Moran et al.(1978)]{moran78} Moran, J.~M., Reid, M.~J., Lada,
  C.~J., Yen, J.~L., Johnston, K.~J., \& Spencer, J.~H.\ 1978, \apj,
  224, L67

\bibitem[Moscadelli et al.(1999)]{moscadelli99} Moscadelli, L.,
  Menten, K.~M., Walmsley, C.~M., \& Reid, M.~J.\ 1999, \apj, 519, 244

\bibitem[Moscadelli et al.(2002)]{moscadelli02} Moscadelli, L.,
  Menten, K.~M., Walmsley, C.~M., \& Reid, M.~J.\ 2002, \apj, 564, 813

\bibitem[Moscadelli et al.(2003)]{moscadelli03} Moscadelli, L.,
  Menten, K.~M., Walmsley, C.~M., \& Reid, M.~J.\ 2003, \apj, 583, 776

\bibitem[Nedoluha \& Watson(1991)]{nedoluha91} Nedoluha, G.~E., \&
  Watson, W.~D.\ 1991, \apj, 367, L63

\bibitem[Norris et al.(1993)]{norris93} Norris, R.~P., Whiteoak,
  J.~B., Caswell, J.~L., Wieringa, M.~H., \& Gough, R.~G.\ 1998, \apj,
  412, 222

\bibitem[Norris et al.(1998)]{norris98} Norris, R.~P., et al.\ 1998,
  \apj, 508, 275

\bibitem[Palmer et al.(2003)]{palmer03} Palmer, P., Goss, W.~M., \&
  Devine, K.~E.\ 2003, \apj, 599, 324

\bibitem[Pavlakis \& Kylafis(1996a)]{pavlakis96a} Pavlakis, K.~G., \&
  Kylafis, N.~D.\ 1996a, \apj, 467, 300

\bibitem[Pavlakis \& Kylafis(1996b)]{pavlakis96b} Pavlakis, K.~G., \&
  Kylafis, N.~D.\ 1996b, \apj, 467, 309

\bibitem[Pavlakis \& Kylafis(2000)]{pavlakis00} Pavlakis, K.~G., \&
  Kylafis, N.~D.\ 2000, \apj, 534, 770

\bibitem[Reid et al.(1980)]{reid80} Reid, M.~J., Haschick, A.~D.,
  Burke, B.~F., Moran, J.~M., Johnston, K.~J., \& Swenson, G.~W., Jr.\
  1980, \apj, 239, 89

\bibitem[Reid \& Moran(1988)]{reid88} Reid, M.~J., \& Moran, J.~M.\
  1988, Galactic and Extragalactic Radio Astronomy, 255

\bibitem[Reid et al.(1987)]{reid87} Reid, M.~J., Myers, P.~C., \&
  Bieging, J.~H.\ 1987, \apj, 312, 830

\bibitem[Sams et al.(1996)]{sams96} Sams, B.~J., III, Moran, J.~M., \&
  Reid, M.~J.\ 1996, \apj, 459, 632

\bibitem[Shu(1992)]{shu92} Shu, F.~H.\ 1992, The Physics of
  Astrophysics v.~2: Gas Dynamics (Sausalito: University Science)

\bibitem[Slysh et al.(2001)]{slysh01} Slysh, V.~I., et al.\ 2001,
  \mnras, 320, 217

\bibitem[Strelnitski et al.(2002)]{strelnitski02} Strelnitski, V.,
  Alexander, J., Gezari, S., Holder, B.~P., Moran, J.~M., \& Reid,
  M.~J.\ 2002, \apj, 581, 1180

\bibitem[Turner \& Welch(1984)]{turner84} Turner, J.~L., \& Welch,
  W.~J.\ 1984, \apj, 287, L81

\bibitem[Vlemmings \& van Langevelde(2005)]{vlemmings05} Vlemmings,
  W.~H.~T., \& van Langevelde, H.~J.\ 2005, \aap, 434, 1021

\bibitem[Watson et al.(2002)]{watson02} Watson, W.~D., Sarma, A.~P.,
  \& Singleton, M.~S.\ 2002, \apj, 570, L37

\bibitem[Watson \& Wyld(2003)]{watson03} Watson, W.~D., \& Wyld,
  H.~W.\ 2003, \apj, 598, 357

\bibitem[Welch \& Marr(1987)]{welch87} Welch, W.~J., \& Marr, J.\
  1987, \apj, 317, L21

\bibitem[Wilner et al.(1999)]{wilner99} Wilner, D.~J., Reid, M.~J., \&
  Menten, K.~M.\ 1999, \apj, 513, 775

\bibitem[Wright et al.(2004a)]{wright04a} Wright, M.~M., Gray, M.~D.,
  \& Diamond, P.~J.\ 2004a, \mnras, 350, 1253

\bibitem[Wright et al.(2004b)]{wright04b} Wright, M.~M., Gray, M.~D.,
  \& Diamond, P.~J.\ 2004b, \mnras, 350, 1272

\bibitem[Xu et al.(2006)]{xu06} Xu, Y., Reid, M.~J., Zheng, X.~W., \&
  Menten, K.~M.\ 2006, Science, 311, 54

\bibitem[Yorke(1986)]{yorke86} Yorke, H.~W.\ 1986, \araa, 24, 49

\end{thebibliography}
\end{document}